\documentclass[preprint,11pt]{article}

\usepackage[utf8]{inputenc}
\usepackage{a4wide}
\usepackage[fleqn]{amsmath}
\usepackage{amssymb}
\usepackage{amsfonts}
\usepackage{mathtools}
\usepackage{amsthm}
\usepackage{graphicx}
\usepackage{slashed}
\usepackage{multirow}
\usepackage{listings}
\usepackage{color}
\usepackage{cite}

\usepackage{booktabs}
\usepackage{array,multirow}
\usepackage{dsfont}
\usepackage{url}
\usepackage{subcaption}
\usepackage{changepage}
\usepackage[table]{xcolor}

\usepackage{fancyhdr}

\usepackage[section]{placeins}
\usepackage{float}
\usepackage{authblk}
\restylefloat{table}

\usepackage{tabularx}
\usepackage{rotating}
\usepackage{etex,etoolbox}
\usepackage{physics}

\DeclareMathOperator{\Li}{Li}

\usepackage{tikz-feynman}
\usepackage{arydshln}
\makeatletter
\def\adl@drawiv#1#2#3{
	\hskip.5\tabcolsep
	\xleaders#3{#2.5\@tempdimb #1{1}#2.5\@tempdimb}
	#2\z@ plus1fil minus1fil\relax
	\hskip.5\tabcolsep}
\newcommand{\cdashlinelr}[1]{
	\noalign{\vskip\aboverulesep
		\global\let\@dashdrawstore\adl@draw
		\global\let\adl@draw\adl@drawiv}
	\cdashline{#1}
	\noalign{\global\let\adl@draw\@dashdrawstore
		\vskip\belowrulesep}}
\makeatother
\usepackage{booktabs}

\usepackage{glossaries}
\newacronym{SM}{SM}{Standard Model}
\newacronym{IBP}{IBP}{integration-by-parts}
\newacronym{EW}{weak}{weak}
\newacronym{LO}{LO}{leading-order}
\newacronym{NLO}{NLO}{next-to-leading-order}
\newacronym{PDF}{PDF}{parton distribution function}

\newcolumntype{L}[1]{>{\raggedright\let\newline\\\arraybackslash\hspace{0pt}}m{#1}}
\newcolumntype{C}[1]{>{\centering\let\newline\\\arraybackslash\hspace{0pt}}m{#1}}
\newcolumntype{R}[1]{>{\raggedleft\let\newline\\\arraybackslash\hspace{0pt}}m{#1}}
\newcolumntype{N}{@{}m{0pt}@{}}

\numberwithin{equation}{section}

\usepackage[colorlinks=true,linkcolor=blue,citecolor=blue]{hyperref}

\usepackage[title,titletoc]{appendix}

\definecolor{shaded}{RGB}{245,245,245}
\usepackage{placeins}

\newcommand{\intd}[1]{\left\langle#1\right\rangle} 
\newcommand{\axl}{\mathrm{A}} 
\newcommand{\vcr}{\mathrm{V}}

\newcommand{\collier}{{\sc\small COLLIER}}

\newcommand{\Fermat}{{\sc\small Fermat}}

\newcommand{\FORM}{{\sc\small FORM}}

\newcommand{\Kira}{{\sc\small Kira}}

\newcommand{\mgamclong}{{\sc\small MadGraph5\_aMC@NLO}}
\newcommand{\MadLoop}{{\sc\small MadLoop}}

\newcommand{\mgamc}{{\sc\small MG5aMC}}
\newcommand{\ninja}{{\sc\small Ninja}}

\newcommand{\oneloop}{{\sc\small OneLOop}}

\newcommand{\python}{{\sc\small Python}}
\newcommand{\QGraf}{{\sc\small QGraf}}

\newcommand\prompt{{\tt MG5\_aMC>}}

\chardef\MyArticleWithColor=\pdfcolorstackinit page direct{0 g}

\title{\textbf{One-loop weak corrections to Higgs production}}
\date{}

\author[1]{Valentin Hirschi}
\author[2]{Simone Lionetti}
\author[3]{Armin Schweitzer}
\affil[1,3]{\emph{\small ETH Z\"urich, 
R\"amistrasse 101, 
8092 Z\"urich, Switzerland}}
\affil[2]{\emph{\small 
Institute for Particle Physics Phenomenology, 
Durham University, 
Durham DH1 3LE, UK}}

\begin{document}

\glsunset{EW}

\maketitle
\thispagestyle{fancy}

\begin{abstract}

We compute mixed QCD-weak corrections to inclusive Higgs production
at the LHC from the partonic process $g g \to H q \bar{q}$.
We start from the UV- and IR-finite one-loop weak amplitude
and consider its interference
with the corresponding one-loop QCD amplitude.
This contribution is a
$\mathcal{O}(\alpha_s\alpha)$
correction to the leading-order gluon-fusion cross section,
and was not numerically assessed in previous works.
We also compute the cross section from the square of this weak amplitude, suppressed by $\mathcal{O}(\alpha^2)$.
Finally, we consider contributions
from the partonic process $g q \to H q$,
which are one order lower in $\alpha_s$,
as a reference for the size of terms which are not enhanced
by the large gluon luminosity.
We find that, given the magnitude of the uncertainties
on current state-of-the-art predictions for Higgs production,
all contributions computed in this work can be safely ignored,
both fully inclusively and in the boosted Higgs regime.
This result supports the approximate factorisation
of QCD and weak corrections to that process.

\end{abstract}

\clearpage

\setcounter{tocdepth}{1}
\tableofcontents

\section{Introduction}
\label{sec:introduction}

In the quest towards an ever more accurate prediction for the inclusive Higgs production cross section at hadron colliders,
one of the major tasks is the computation of fixed-order corrections in the perturbative expansion in powers of the \gls{SM} couplings.
Our understanding of pure QCD corrections, which are known to be very important for this process, has reached an unprecedented level of accuracy in recent times.
A milestone in this programme was achieved with the computation of the third correction term in the expansion in the strong coupling $\alpha_s$ of the cross section for Higgs production via gluon fusion in the infinite top mass limit \cite{Anastasiou:2015ema,Mistlberger:2018etf}.
In a typical setup for the LHC running at a centre-of-mass energy of 13 TeV,
this contribution shifts the prediction for the total cross section upwards by roughly 3\% \cite{Anastasiou:2016cez}.

On the other hand, \gls{EW} corrections to the \gls{LO} inclusive Higgs cross section also need to be considered.
In the same setup mentioned before, the first \gls{EW} term turns out to increase the total gluon fusion cross section by a significant 5\%~\cite{Aglietti:2004nj,Actis:2008ug,Actis:2008ts}.
Since \gls{NLO} QCD corrections can be as large as the leading contribution,
the motivation to investigate mixed first-order QCD and first-order \gls{EW} corrections is very strong.
Although the exact size of this term is at present unknown,
various approximations have been considered in the literature.
The first estimate to appear was based on the argument
that mixed QCD-\gls{EW} effects on the inclusive Higgs production cross section
are well approximated by combining the purely \gls{EW} term
and the full QCD series in a multiplicative fashion~\cite{Anastasiou:2008tj}.
Following this \emph{factorisation} approach, the authors of ref.~\cite{Anastasiou:2016cez} reported the mixed QCD-\gls{EW} corrections to be approximately 3\% of the full result,
and conservatively estimated the uncertainty stemming from non-factorisable contributions to be 1\% of the total.  The estimates of \cite{Anastasiou:2008tj,Anastasiou:2016cez} are obtained by considering the unphysical limit $m_H\ll m_W,m_Z$. The gluon induced interference contributions discussed in our work are suppressed in this limit by two powers of the weak boson masses with respect to the leading order $\mathcal{O}(\alpha_{S}^2\alpha)$  cross section, which we verified by explicit calculation.
The theoretical uncertainty associated to each of the other main error sources
(determination of \acrlongpl{PDF}, truncation of the QCD perturbative series, and missing quark-mass effects)
is currently of the same order.
It is therefore highly desirable to remove the ambiguity due to the factorisation approximation.

Important steps have recently been made in this direction.
Thanks to the calculation of the three-loop mixed QCD-\gls{EW} correction
to Higgs boson gluon fusion for arbitrary masses
of the $W$, $Z$, and Higgs bosons~\cite{Bonetti:2017ovy},
an estimate of the cross section in the soft-virtual approximation
was obtained~\cite{Bonetti:2018ukf}.
An independent work considered three-loop matrix elements
in the limit of massless vector bosons instead,
and combined them with a different class
of two-loop real-emission contributions~\cite{Anastasiou:2018adr}.
The estimates obtained using these approximations support
the validity of the factorisation approach,
since they include some non-factorisable effects
and find that these are numerically small.

In order for the full mixed QCD-\gls{EW} term to become available, however,
two pieces of the puzzle are still missing.
On the one hand there is the formidable challenge
of computing two-loop matrix elements with an extra real emission
for arbitrary $W$, $Z$, and Higgs masses.
On the other hand, there are UV- and IR-finite one-loop \gls{EW} contributions to the production of the Higgs in association with two partons,
which feature more complicated kinematics but whose one-loop integrals are well understood.
Although in general corrections with fewer or soft real emissions
are expected to dominate within the inclusive cross section~\cite{Bonetti:2018ukf},
the contributions with two extra hard partons are formally of the same order
and may disrupt the approximate factorisation of \gls{EW} and QCD corrections
because of their final-state kinematic structure.

In the present paper, we address this issue by carrying out the exact \emph{inclusive} computation
of the contribution to mixed QCD-\gls{EW} corrections from the one-loop partonic subprocess $g g \to H q \bar{q}$.
We stress that this contribution features one-loop pentagon topologies
which appear only in matrix elements with (at least) two real emissions,
that do not fit in a factorised picture and that have not been assessed before.

The paper proceeds as follows.
In Section~\ref{sec:contributions},
we discuss the different contributions that enter our computation,
we categorise them and identify potential competing mechanisms which are formally of the same order or slightly higher.
Although the computation of the required matrix elements is straightforward using standard public codes for one-loop calculations,
the computation of the pieces of cross sections we are interested in
requires the renormalisation of parton distributions
and the subtraction of initial-state collinear singularities.
Given the very special features of the process examined,
these steps require some care
and are thus described in Section~\ref{sec:singularities}.
Finally, we report and discuss numerical results.

\section{Classification of contributions}
\label{sec:contributions}

In order to classify contributions to the Higgs inclusive cross section, it is useful to write its mixed QCD and \gls{EW} expansion as
\begin{equation}
	\sigma_{pp\to H+X} = \sum_{m,n} \sigma^{(m,n)}_{pp\to H+X},
	\qquad\text{where}\qquad
	\sigma^{(m,n)}_{pp\to H+X} \propto \alpha_s^{m+2} \alpha^{n+1},
\end{equation}
where the prefactor $\alpha_s^2\alpha^1$ is chosen so as to match the couplings factorised by the leading-order loop-induced gluon-fusion contribution to inclusive Higgs production.
Notice that we group all squared couplings that are not strong, including the Yukawa of the top quark, under the label $\alpha$, in view of their comparable strength and of the electroweak gauge relations often rendering their separate factorisation ambiguous.
The corrections often labelled ``QCD N$^m$LO'' and ``(electro)weak N$^n$LO'' are then denoted by $\sigma^{(m,0)}_{pp\to H+X}$ and $\sigma^{(0,n)}_{pp\to H+X}$,
as they become impractical when addressing the mixed cases $\sigma^{(m,n)}_{pp\to H+X}$.
With such a notation in mind,
the expected naive parametric suppression from the couplings,
which counts $\alpha_s\sim 10^{-1}$ and $\alpha\sim 10^{-2}$,
simply reads $\sigma^{(m,n)}_{pp\to H+X}\sim10^{-m-2n}$.
In order to discuss interference terms,
we also find it useful to introduce a similar notation for amplitudes:
\begin{equation}
	A^{(i,j)}_{a b\to H+X} \propto g_s^{i+2} g^{j+1},
\end{equation}
where we denote by $g$ all couplings that are not $g_s$.

As mentioned above,
\gls{EW} and QCD corrections are expected to factorise to a certain degree,
such that
\begin{equation}
	\sigma^{(m,n)}_{pp\to H+X} \sim
	\sigma^{(m,0)}_{pp\to H+X} \cdot \sigma^{(0,n)}_{pp\to H+X}.
\end{equation}
This approximation is valid under the assumption that the main contributions
to the mixed QCD-\gls{EW} cross section are to be attributed
either to soft gluons or Sudakov \gls{EW} logarithms.
If one is to assess violations of this factorisation,
the expansion term $\sigma^{(1,1)}_{pp\to H+X}$ must be computed \emph{exactly}.
We now set out to discuss the many contributions this term receives.

In this work, we only consider \emph{weak} corrections involving the $W$ and $Z$ bosons, as these dominate over the genuine electroweak corrections (\emph{i.e.} unresolved \emph{photon} exchange or emission) to contributions where the Higgs is produced from massive quark loop lines that are not the top-quark.

Also, the gluon initiated processes are expected to be the dominant contributions at the LHC, where quark \glspl{PDF} are small in comparison to the gluon one for the typical values of the Bjorken $x$'s probed by the kinematics involved.
We therefore neglect all contributions to $\sigma^{(1,1)}_{pp\to H+X}$ that factorise parton luminosities with at least one quark.
To get a reference for the size of these terms that we do not compute,
we report numerical results also for $\sigma^{(0,1)}_{gq\to Hq}$.
\footnote{Note that our initial-state notation $g q$ encompasses in this context both permutations $g q$ and $q g$.}

Weak corrections stemming from the interference with leading QCD production modes are often subject to kinematic suppressions that renders them smaller than what is naively expected from their factorised couplings.
For this reason, we also report the pieces of the cross sections $\sigma^{(0,2)}_{gg\to H q \bar{q}}$ and $\sigma^{(-1,2)}_{gq\to H q}$
built from the square of the amplitudes $A^{(0,2)}_{gg\to H q \bar{q}}$ and $A^{(-1,2)}_{gq\to H q}$.
These form a gauge-invariant subset of higher-order contributions.

Our work reports on the contribution $\sigma^{(1,1)}_{gg\to Hq \bar{q}}$ for the first time and, together with the results from refs.~\cite{Anastasiou:2008tj,Bonetti:2018ukf}, it completes the computation of $\sigma^{(1,1)}_{g g \to H + X}$.
We now proceed to list in Table~\ref{InterferencesTable} all amplitudes building $\sigma^{(1,1)}_{gp\to H+X}$.

\renewcommand{\arraystretch}{1.5}
\begin{table}[ht!]
\begin{center}
{\setlength\doublerulesep{1.5pt}   
	\aboverulesep=0.1ex 
	\belowrulesep=-0.11ex
\begin{tabular}{c||c|c|c|c||c|c||c|c}
$\cross$ 						&$ A^{(0,0)\star}_{gg\to H} $ 		& $ A^{(2,0)\star}_{gg\to H} $ 		&$ A^{(0,2)\star}_{gg\to H} $ 		& $ A^{(2,2)\star}_{gg\to H} $ & $ A^{(1,0)\star}_{gg\to Hg} $ 		& $ A^{(1,2)\star}_{gg\to Hg} $ 		& $A^{(2,0)\star}_{gg\to H q \bar{q}} $ & $A^{(0,2)\star}_{gg\to H q \bar{q}}$ \\
\hline
$ A^{(0,2)}_{gg\to H} $	 		& \cellcolor{blue!25} $\sigma^{(0,1)}_{gg\to H}$ 	& \fbox{\cellcolor{blue!25}$\sigma^{(1,1)}_{gg\to H}$}	& $\sigma^{(0,2)}_{gg\to H}$	& $\sigma^{(1,2)}_{gg\to H}$
							&\multicolumn{4}{c}{} \\
$ A^{(2,2)}_{gg\to H} $ 			& \fbox{\cellcolor{blue!25}$\sigma^{(1,1)}_{gg\to H}$} 	& $\sigma^{(2,1)}_{gg\to H}$	& $\sigma^{(1,2)}_{gg\to H}$	& $\sigma^{(2,2)}_{gg\to H}$
							&\multicolumn{4}{c}{}  \\ \hline
$ A^{(1,2)}_{gg\to H g} $ 			&\multicolumn{3}{c}{} & 	
							& \fbox{\cellcolor{blue!25}$\sigma^{(1,1)}_{gg\to Hg}$}	& $\sigma^{(1,2)}_{gg\to Hg}$	& \multicolumn{2}{c}{} \\ \hline
$A^{(0,2)}_{gg\to H q \bar{q}}$	 	&\multicolumn{5}{c}{} &
							& \fbox{\cellcolor{green!45}$\sigma^{(1,1)}_{gg\to Hq \bar{q}}$}	& \cellcolor{green!45} $\sigma^{(0,2)}_{gg\to Hq \bar{q}}$	\\
\end{tabular}
\begin{tabular}{c||c|c|c|c}
$\cross$ & 
$ A^{(1,0)\star}_{gq\to H q} $ & $ A^{(3,0)\star}_{gq\to H q} $ & $ A^{(-1,2)\star}_{gq\to H q} $ & $ A^{(1,2)\star}_{gq\to H q} $  \\
\hline
$ A^{(-1,2)}_{gq\to H q} $		&  \cellcolor{green!45} $\sigma^{(0,1)}_{gq\to Hq}$ & \fbox{$\sigma^{(1,1)}_{gq\to Hq}$} &   \cellcolor{green!45} $\sigma^{(-1,2)}_{gq\to Hq}$ & $\sigma^{(0,2)}_{gq\to Hq}$ \\
$ A^{(1,2)}_{gq\to H q} $ 		& \fbox{$\sigma^{(1,1)}_{gq\to Hq}$} & $\sigma^{(2,1)}_{gq\to Hq}$ & & $\sigma^{(1,2)}_{gq\to Hq}$  \\ 
\end{tabular}
}
\end{center}
\caption{\label{InterferencesTable}Summary of contributing amplitudes to the weak corrections to Higgs inclusive production involving one (bottom table) and two (top table) initial-state gluons, for various perturbative orders. The results reported in this work are highlighted with a green background, while those addressed in ref.~\cite{Anastasiou:2008tj,Bonetti:2018ukf} are denoted in blue. Together, these form the \emph{complete} $\sigma^{(1,1)}_{g g \to H + X}$ \gls{EW} correction. }
\end{table}

We now turn to discussing the Feynman diagrams building the amplitudes $A^{(2,0)}_{gg\to H q \bar{q}} $, $A^{(0,2)}_{gg\to H q \bar{q}}$, $ A^{(1,0)}_{gq\to H q} $ and $ A^{(-1,2)}_{gq\to H q} $ that contribute to the cross sections presented in this work.

\begin{figure}[h]
	\centering
	\begin{subfigure}[b]{.24\linewidth}
		\includegraphics[width=\linewidth]{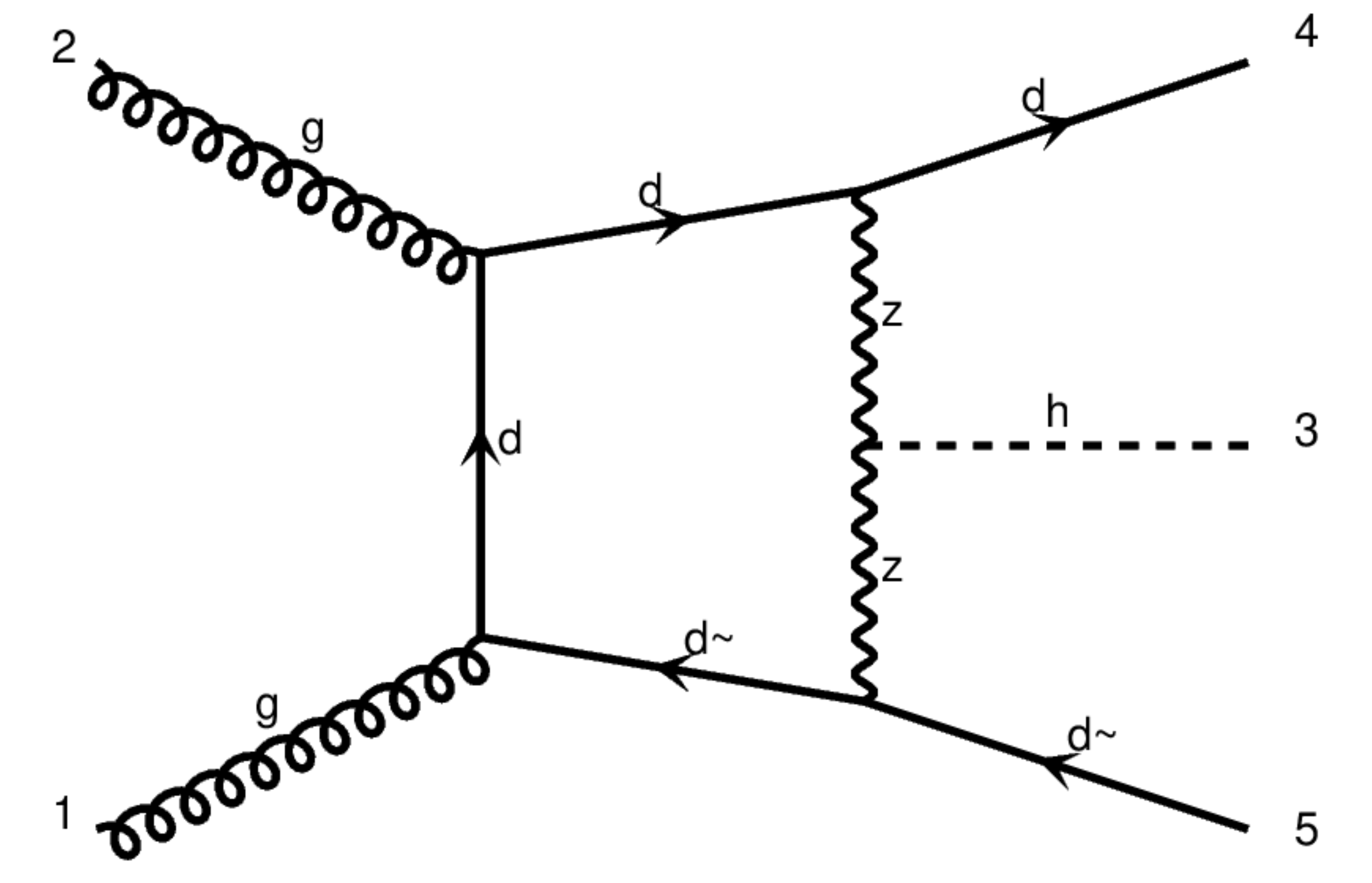}
		\caption{}\label{subfig:pentagon_ggHddb}
	\end{subfigure}
	\begin{subfigure}[b]{.24\linewidth}
		\includegraphics[width=\linewidth]{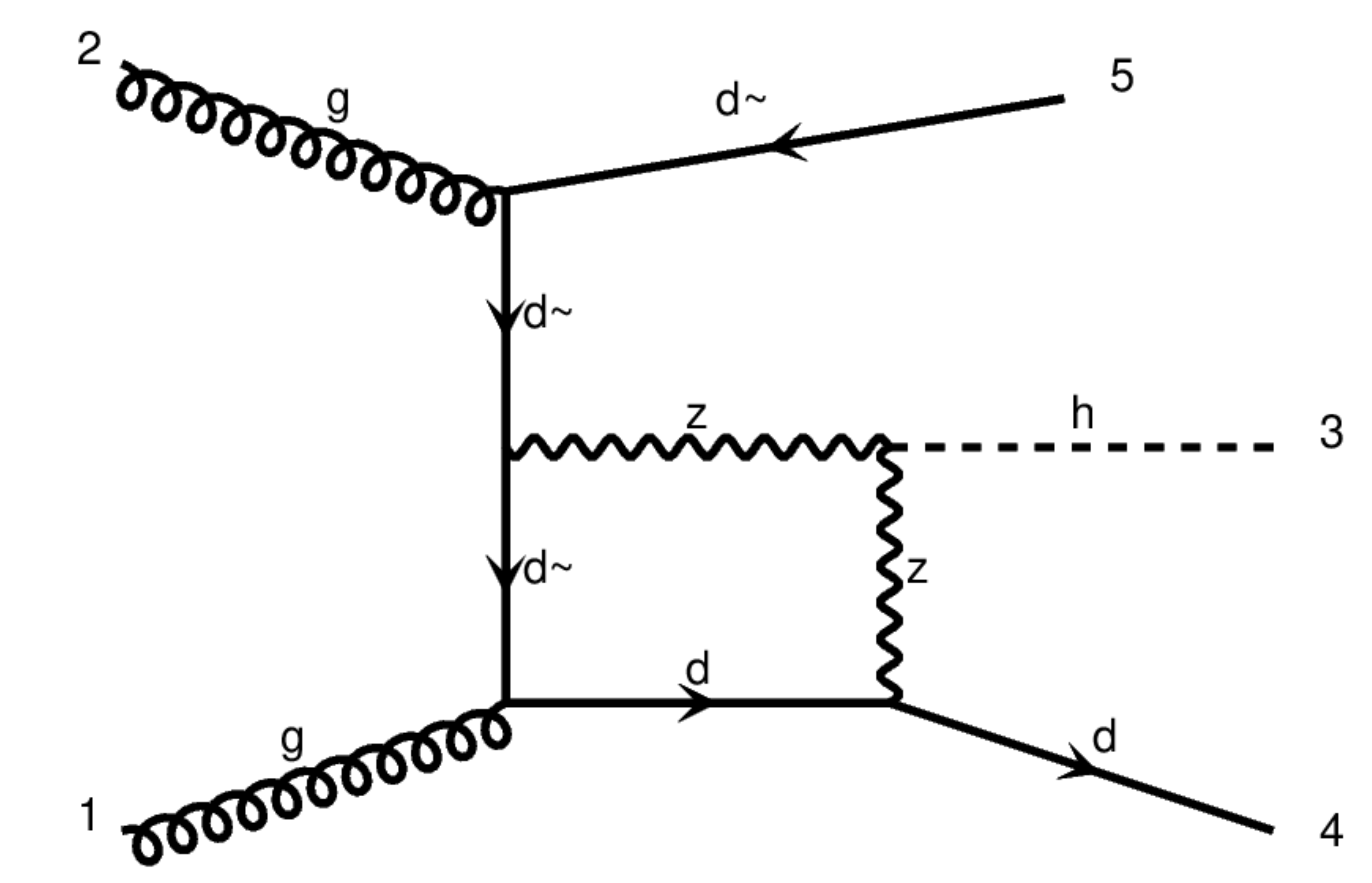}
		\caption{}\label{subfig:box_div_ggHddb}
	\end{subfigure}
	\begin{subfigure}[b]{.2225\linewidth}
		\includegraphics[width=\linewidth]{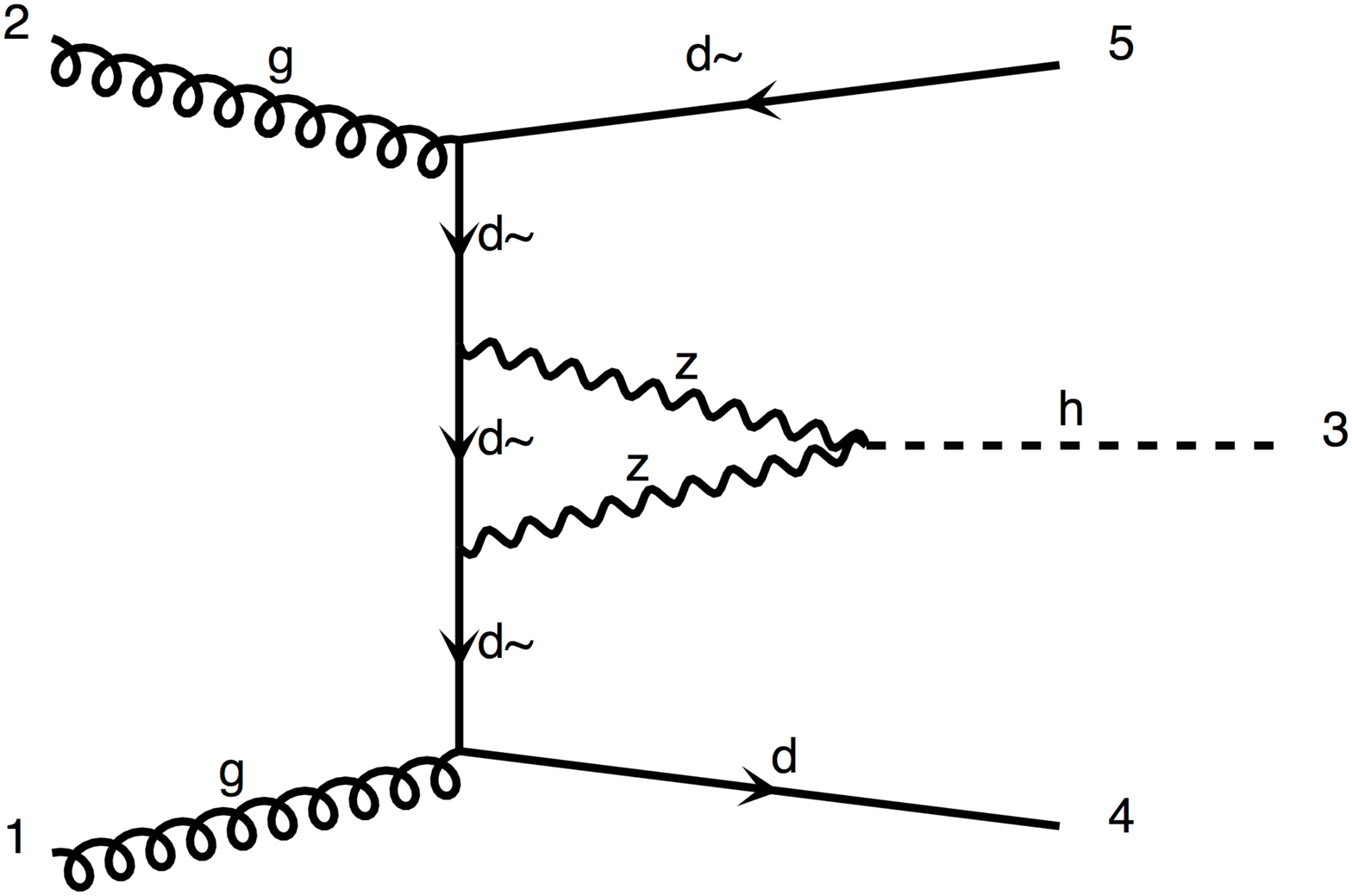}
		\caption{}\label{subfig:triangle_I_ggHddb}
	\end{subfigure}
\\
	\begin{subfigure}[b]{.24\linewidth}
		\includegraphics[width=\linewidth]{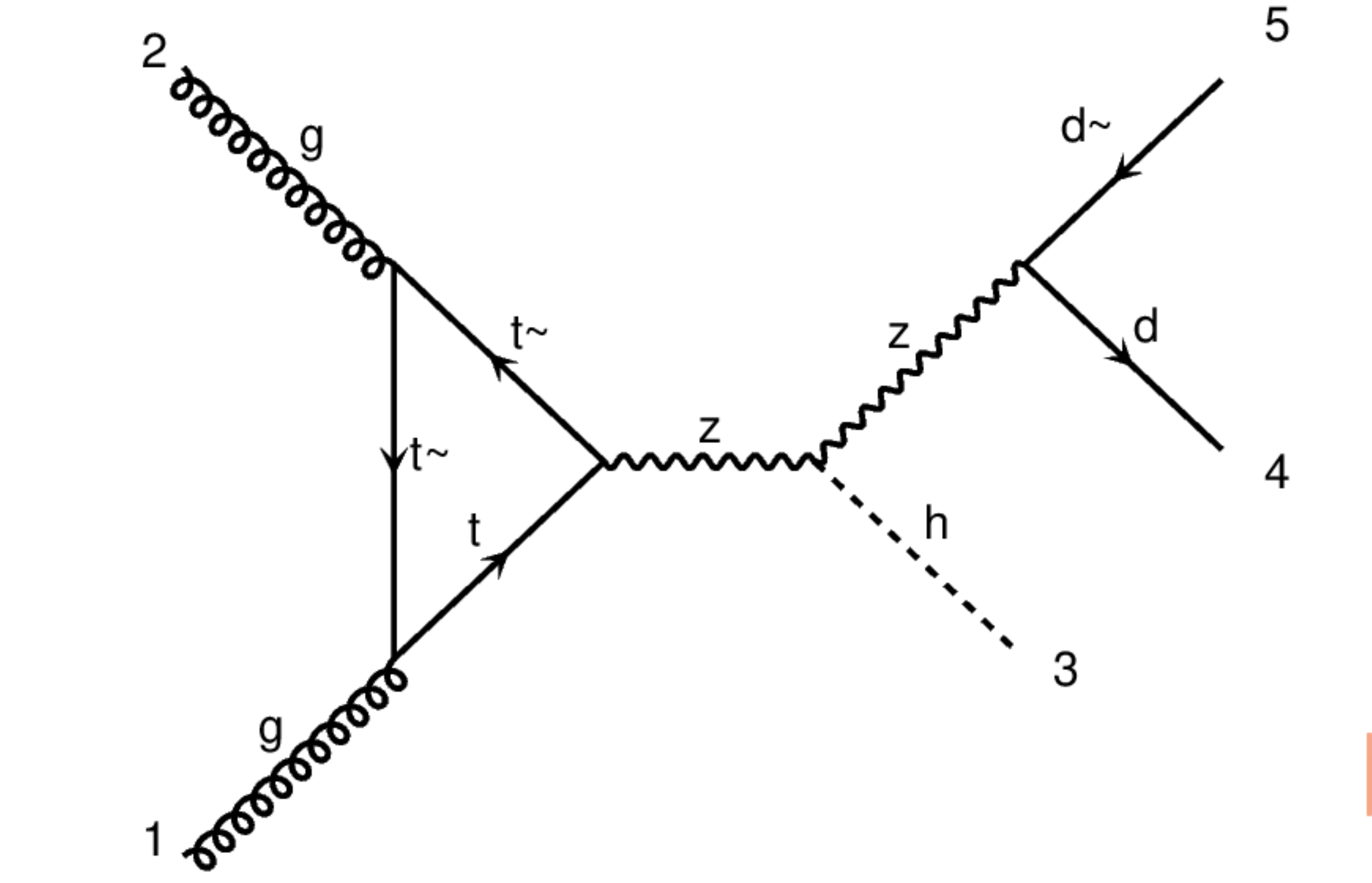}
		\caption{}\label{subfig:triang_associated_prod_ggHddb}
	\end{subfigure}
	\begin{subfigure}[b]{.24\linewidth}
		\includegraphics[width=\linewidth]{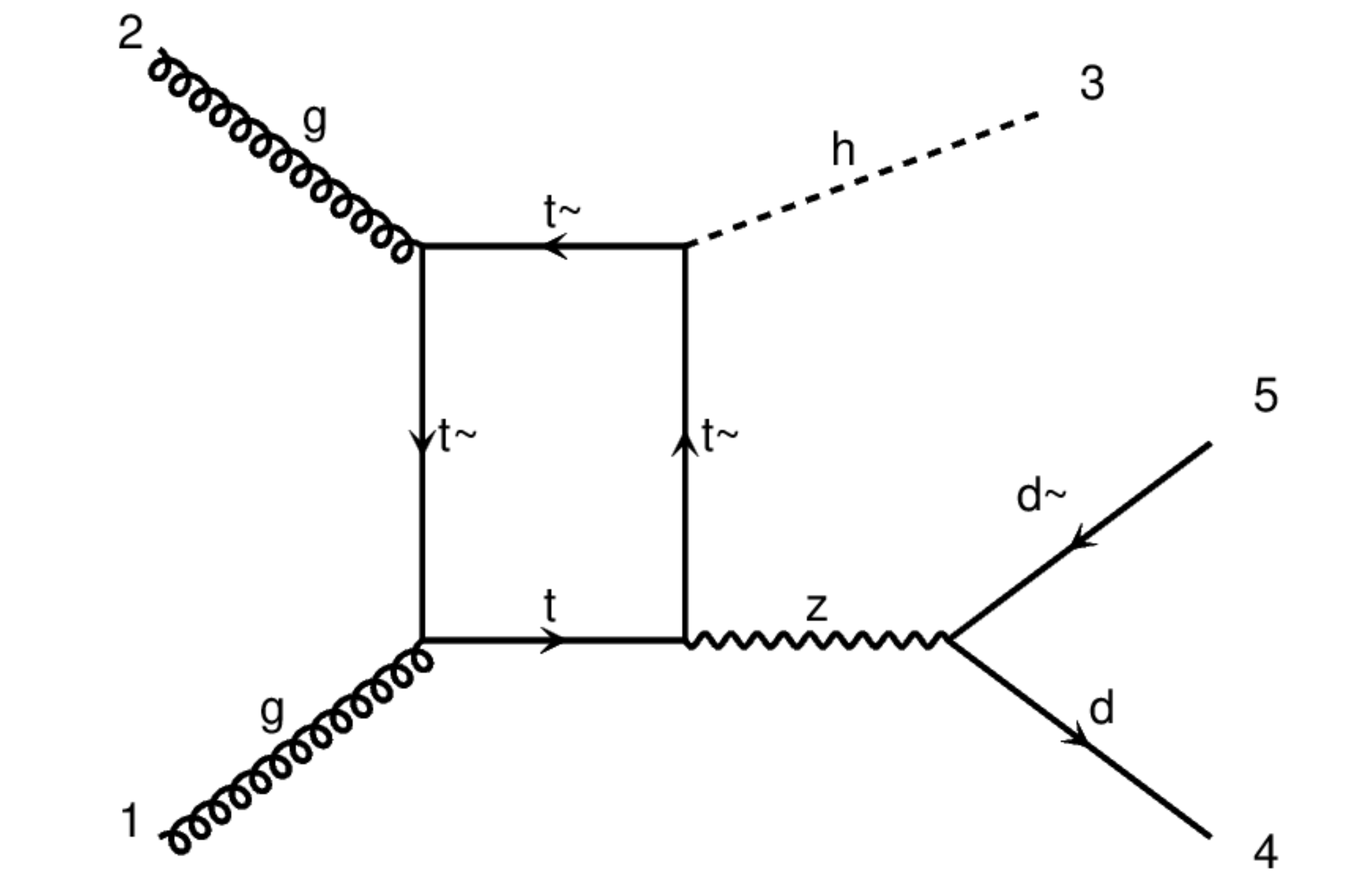}
		\caption{}\label{subfig:box_associated_prod_ggHddb}
	\end{subfigure}
\begin{subfigure}[b]{.24\linewidth}
\includegraphics[width=\linewidth]{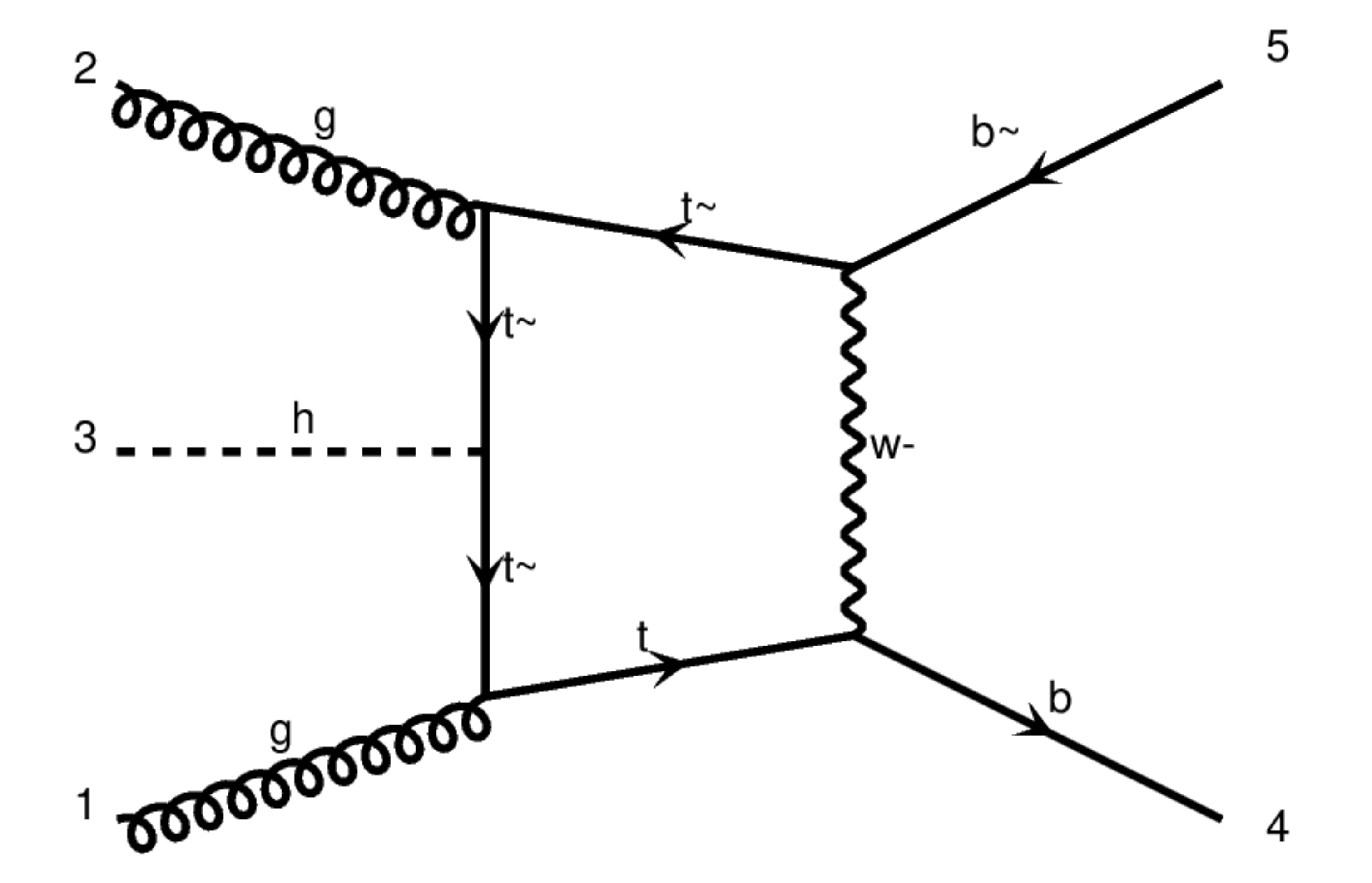}
\caption{}\label{diagF}
\end{subfigure}
	\caption{
	\label{allEWdiagrams}
	Representative subset of diagrams contributing to the amplitude $A^{(0,2)}_{gg\to H q \bar{q}}$.
	In diagrams~\ref{subfig:pentagon_ggHddb},~\ref{subfig:box_div_ggHddb} and~\ref{subfig:triangle_I_ggHddb}, the $Z$ boson can be interchanged with a $W$ boson.
	Diagrams~\ref{subfig:triang_associated_prod_ggHddb} and~\ref{subfig:box_associated_prod_ggHddb} are contributions to the production of a Higgs in association with a $Z$ boson and are only included in the computation of $\sigma^{(1,1)}_{pp\to H+X}$, and not that of $\sigma^{(1,2)}_{pp\to H+X}$. Diagrams of the class~\ref{diagF} are only present for the process $g g \to H b \bar{b}$.
	}
\end{figure}

\begin{figure}[h!]
	\centering
	\begin{subfigure}[b]{.24\linewidth}
		\includegraphics[width=\linewidth]{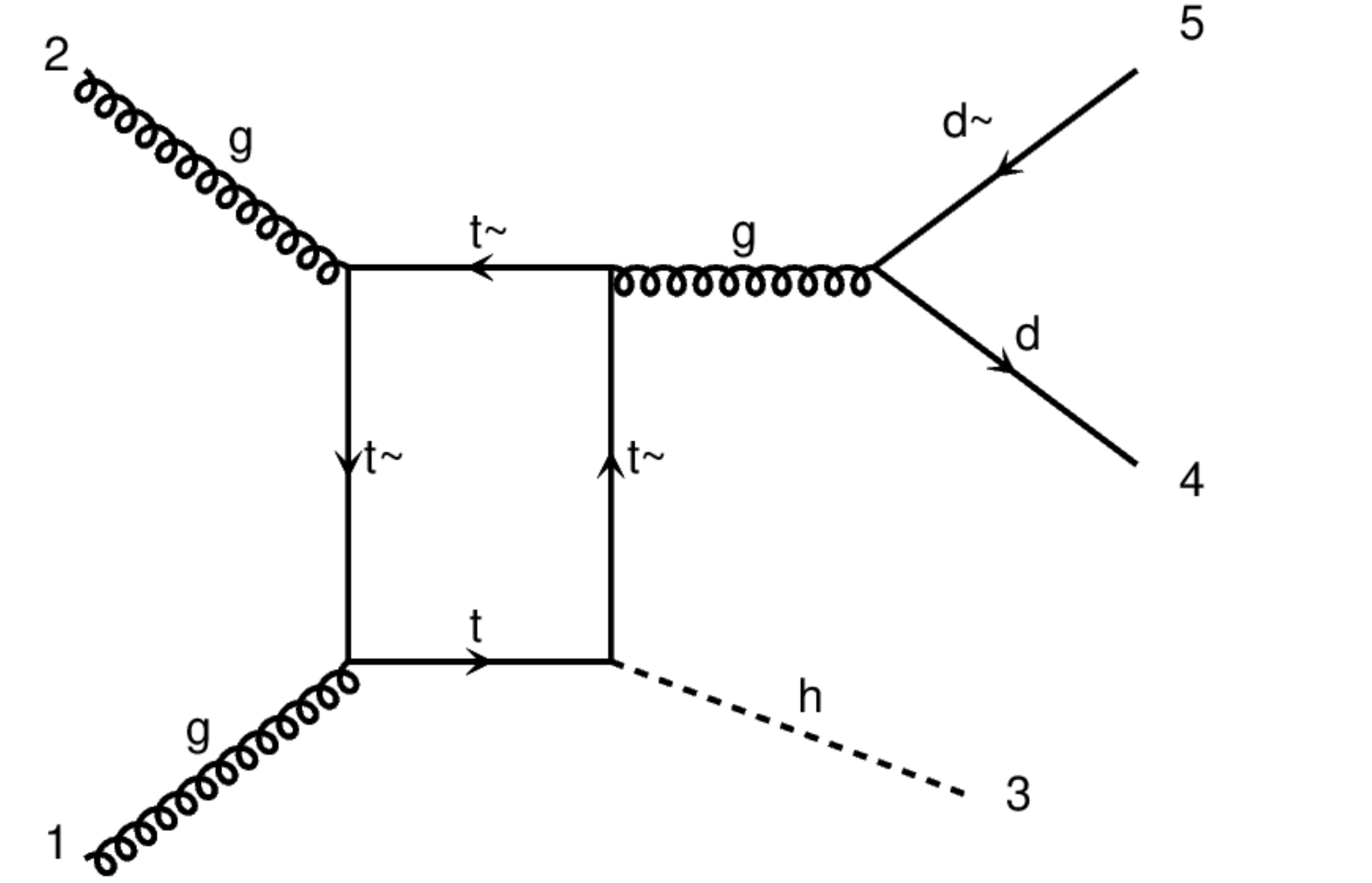}
		\caption{}\label{subfig:box_qcdBG_ggHddb}
	\end{subfigure}
	\begin{subfigure}[b]{.24\linewidth}
		\includegraphics[width=\linewidth]{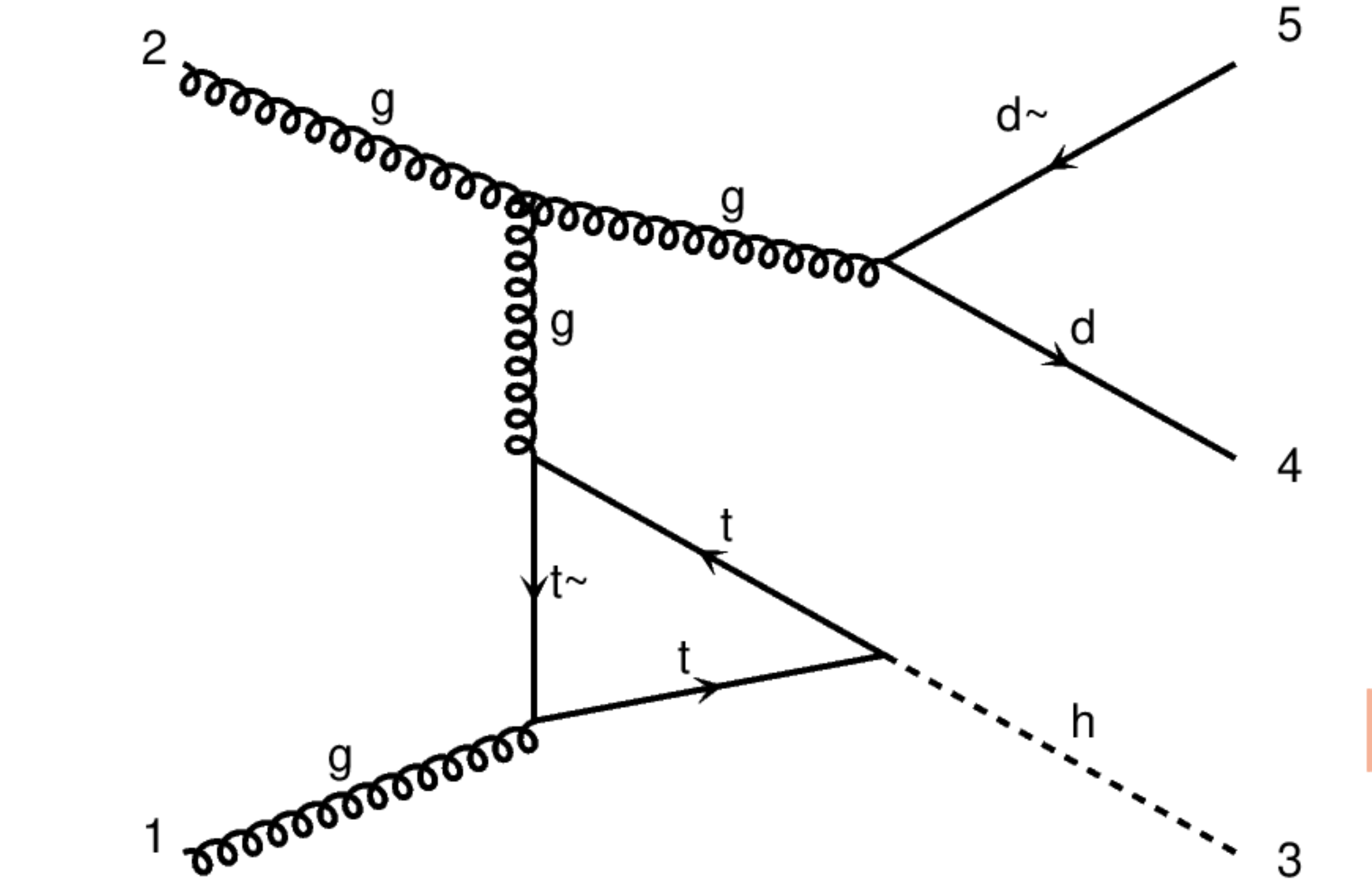}
		\caption{}\label{subfig:triangle_qcdBG_ggHddb}
	\end{subfigure}
	\begin{subfigure}[b]{.24\linewidth}
	\includegraphics[width=\linewidth]{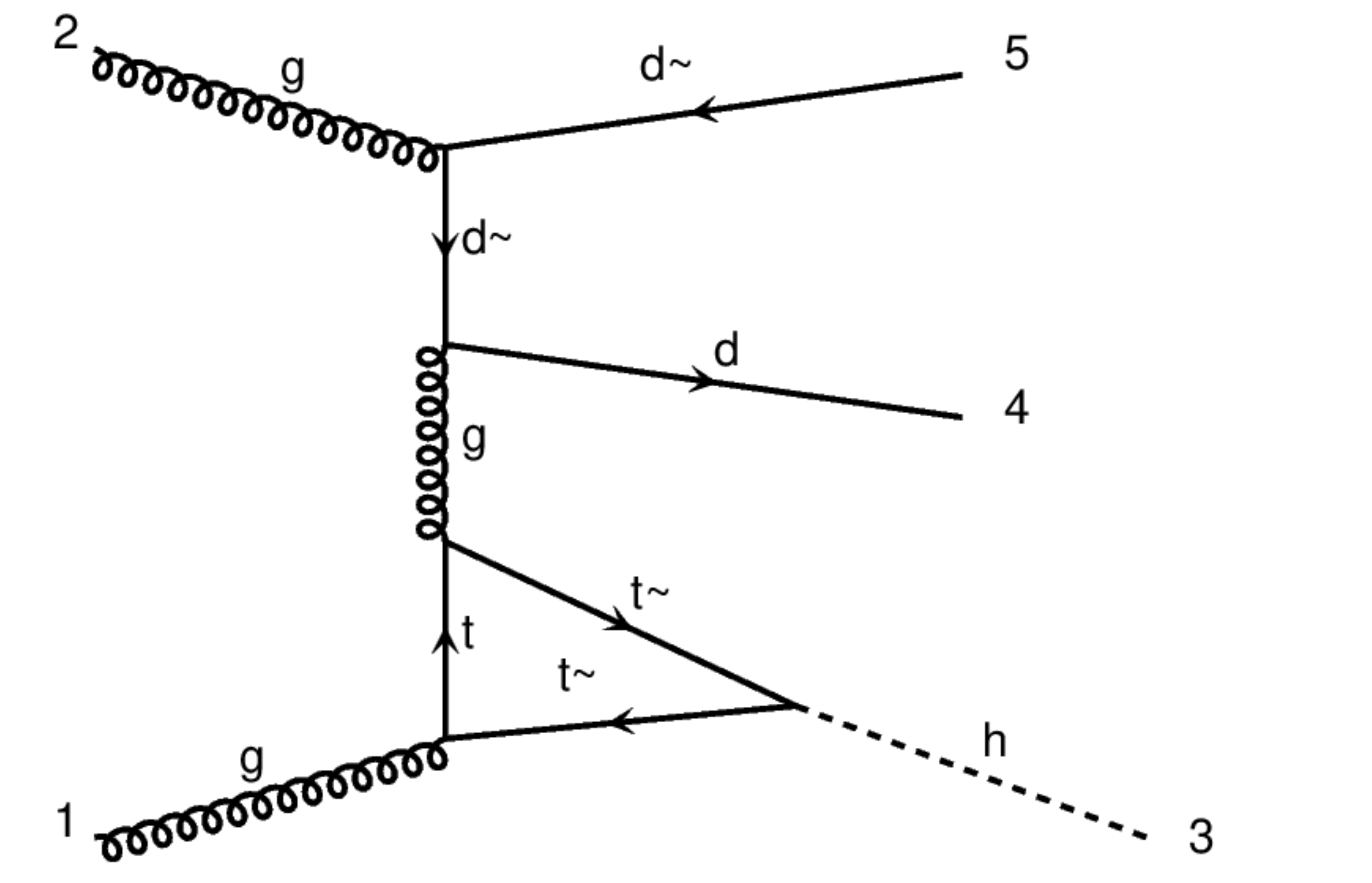}
	\caption{}\label{subfig:triangle_II_qcdBG_ggHddb}
	\end{subfigure}
	\begin{subfigure}[b]{.24\linewidth}
		\includegraphics[width=\linewidth]{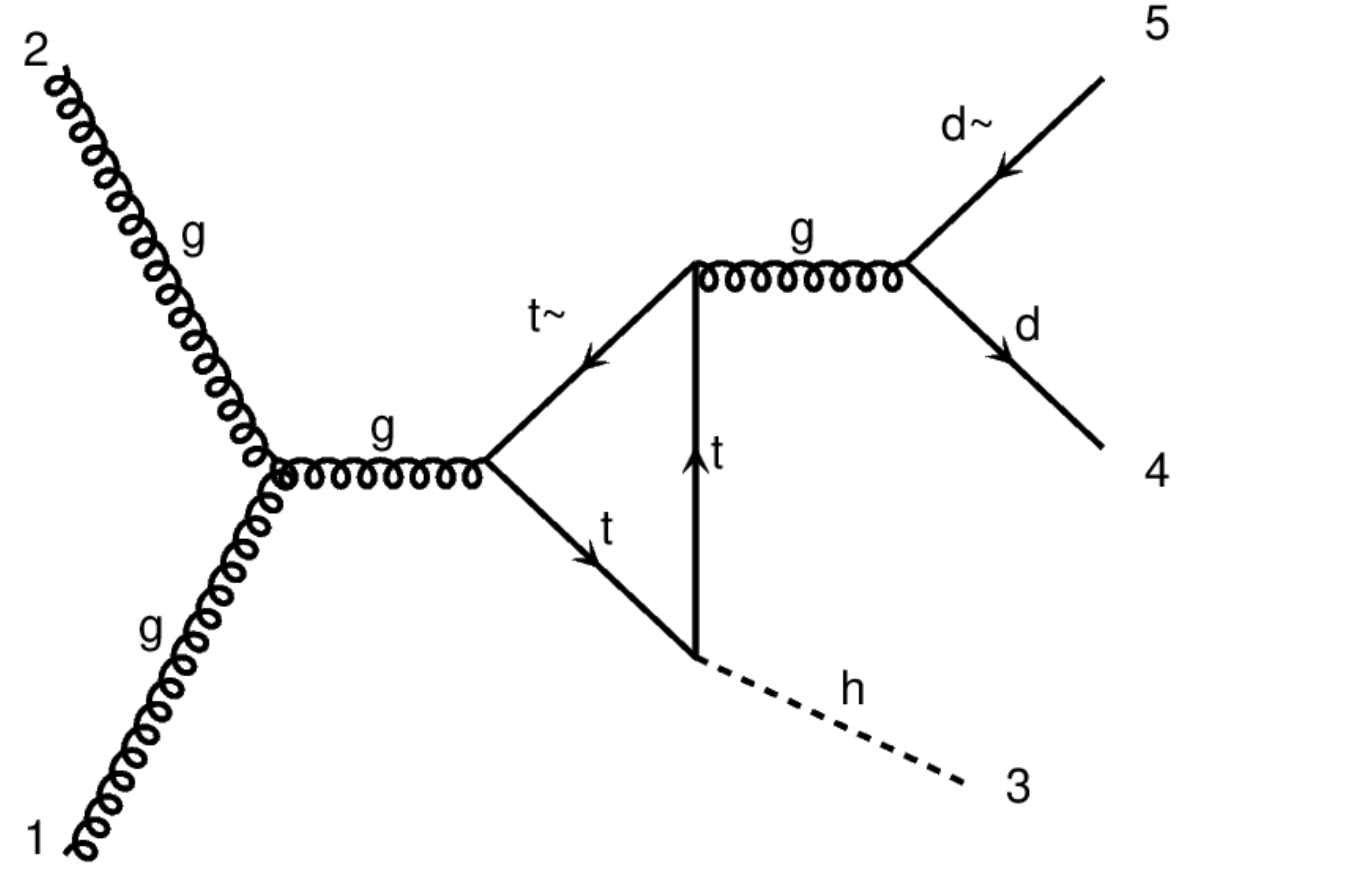}
		\caption{}\label{subfig:triangle_III_qcdBG_ggHddb}
	\end{subfigure}
	\caption{\label{allQCDdiagrams} Diagrams building the amplitude $A^{(2,0)}_{gg\to H q \bar{q}}$ against which the diagrams listed in Fig.~\ref{allEWdiagrams} are interfered to yield $\sigma^{(1,1)}_{gg\to H q \bar{q}}$. The full top-quark mass dependence is retained.
	}
\end{figure}

The amplitude $A^{(2,0)}_{gg\to H q \bar{q}} $ is built from the diagrams depicted in Fig.~\ref{allEWdiagrams} where the Higgs is produced via weak vector boson fusion and interfered with the leading QCD gluon-fusion diagrams shown in Fig.~\ref{allQCDdiagrams}.

Diagrams of the class~\ref{subfig:triang_associated_prod_ggHddb} and~\ref{subfig:box_associated_prod_ggHddb}, where the Higgs is produced via gluon-fusion, feature a Z-boson propagator\footnote{The diagram analogous to~\ref{subfig:box_associated_prod_ggHddb} with a photon instead of the Z-boson is exactly zero in virtue of Furry's theorem.} which however does not yield any Breit-Wigner resonance as they are interfered against the QCD diagrams of Fig.~\ref{allQCDdiagrams}.
We must nonetheless regulate the Z-boson propagator pole, which motivates our use in this computation of the complex-mass scheme~\cite{Denner:1991kt,Frederix:2018nkq} with finite widths for the internal top quark and unstable weak gauge bosons.
These diagrams~\ref{subfig:triang_associated_prod_ggHddb} and~\ref{subfig:box_associated_prod_ggHddb} are however ignored when considering their squared contribution to $\sigma^{(1,1)}_{pp\to H+X}$,
 since in this case they are best accounted for in the narrow-width approximation as the \gls{LO} prediction for associated Higgs production, \emph{i.e.} $\sigma^{(1,1)}_{gg\to HZ}$ (also reported in this work).

Finally, diagrams of the class~\ref{diagF} are specific to the third-generation quarks where the Higgs can also be emitted from the top-quark running in the loop.
This contribution is analogous to that of the heavy quarks in the two-loop electroweak corrections to Higgs production investigated in ref.~\cite{Degrassi:2004mx} and, for this reason, we found it interesting to report our results separately for the processes $g g \to H q \bar{q}$, with $q \equiv u,d,c,s$, and $g g \to b \bar{b} H$.

\begin{figure}[h!]
	\centering
	\begin{subfigure}[b]{.23\linewidth}
		\includegraphics[width=\linewidth]{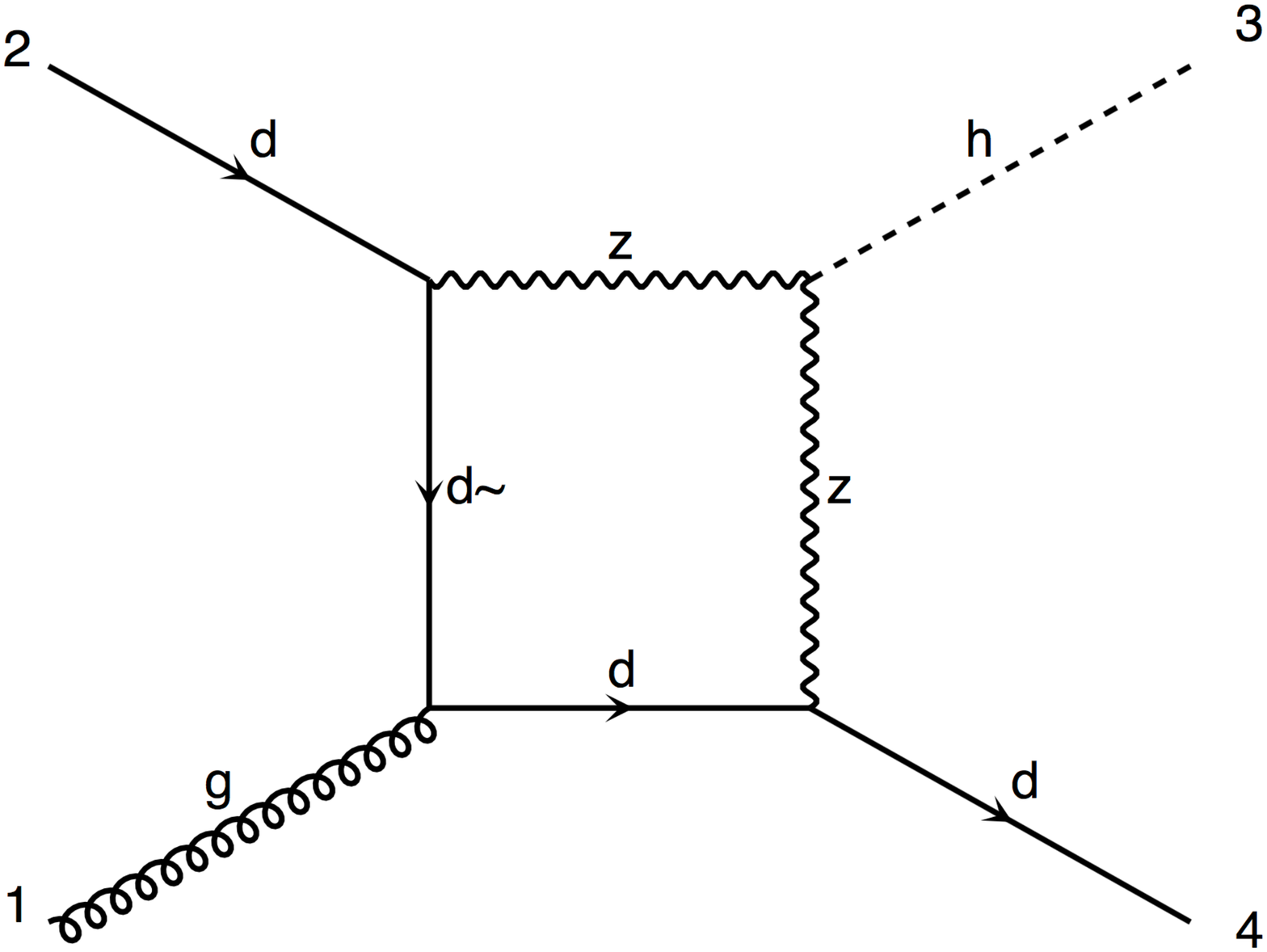}
		\caption{}\label{diagAewborn}
	\end{subfigure}
	\begin{subfigure}[b]{.23\linewidth}
		\includegraphics[width=\linewidth]{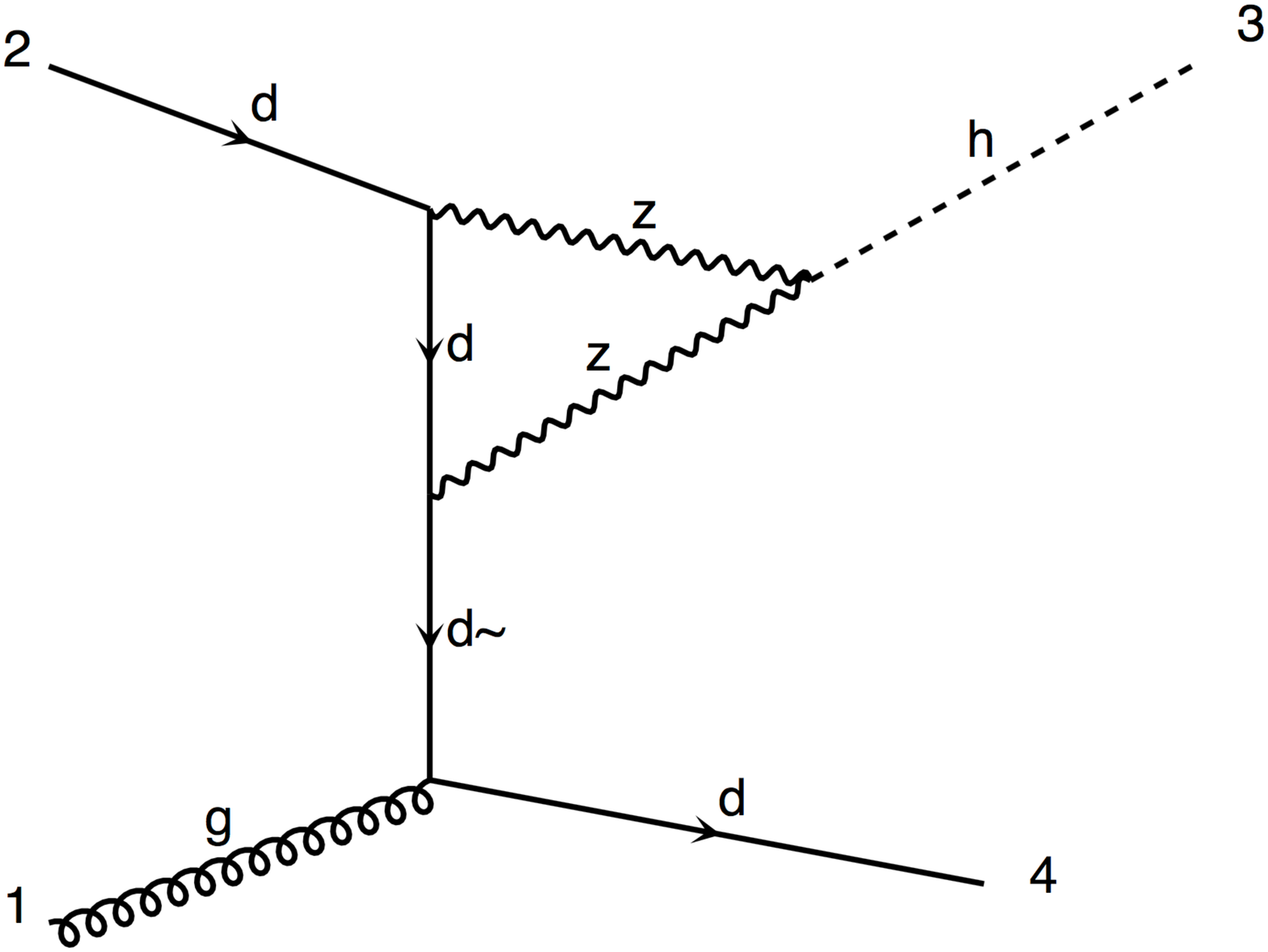}
		\caption{}\label{diagBewborn}
	\end{subfigure}
	\begin{subfigure}[b]{.23\linewidth}
		\includegraphics[width=\linewidth]{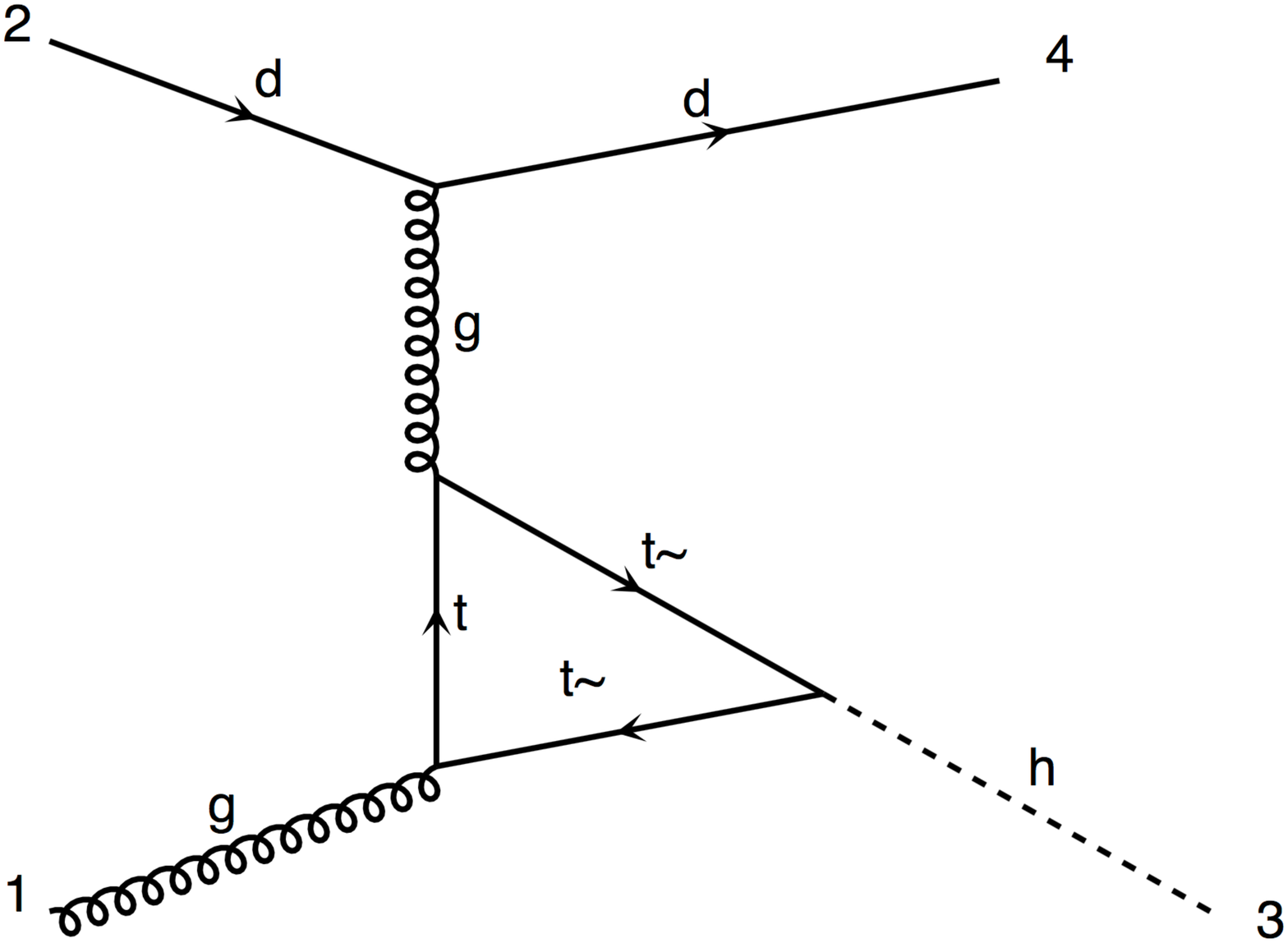}
		\caption{}\label{diagCqcdborn}
	\end{subfigure}
	\begin{subfigure}[b]{.23\linewidth}
		\includegraphics[width=\linewidth]{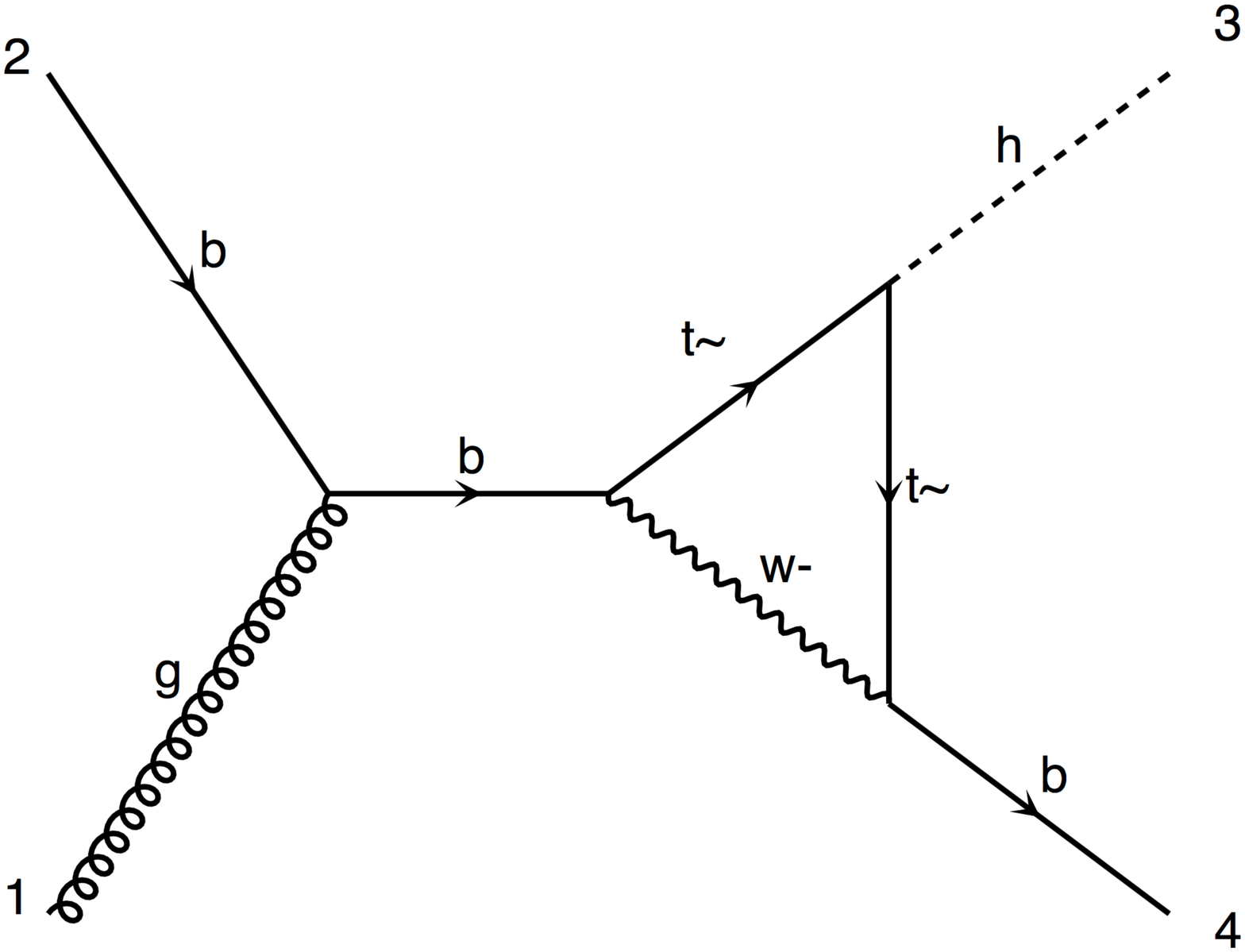}
		\caption{}\label{diagDewborn}
	\end{subfigure}
	\caption{\label{allBorndiagrams} Diagrammatic contributions to the amplitudes $A^{(-1,2)}_{gq\to H q}$ (Figs.~\ref{diagAewborn},~\ref{diagBewborn},~\ref{diagDewborn}) and  $A^{(1,0)}_{gq\to H q}$ (Fig.~\ref{diagCqcdborn}), yielding $\sigma^{(-1,2)}_{gq\to H q}$ and $\sigma^{(0,2)}_{gq\to H q}$ respectively. Diagrams~\ref{diagAewborn} and~\ref{diagBewborn} also appear in the reduced matrix elements factorised by the collinear subtraction local counterterms of Eqs.~\ref{intcoll} and~\ref{intreal}. Notice that diagrams belonging to the class~\ref{diagDewborn} are specific to the process ${g b \to H b}$. In all cases, the full top-quark mass dependence is retained.
	}
\end{figure}

\section{Initial-state collinear singularities}
\label{sec:singularities}

All of the one-loop amplitudes considered in this paper
are free of explicit ultraviolet and infrared divergences
that can arise from the integration over the loop momenta.
In other words,
working in dimensional regularisation with $D \equiv 4-2\epsilon$,
their analytic expressions do not contain explicit poles
in the dimensional regulator $\epsilon$.
However, 
matrix elements may feature non-integrable infrared divergences
in regions of the phase space which correspond to unresolved configurations.
In order to discuss this issue, we concentrate on the amplitude
$A^{(0,2)}_{gg\to H q \bar{q}}$ as it constitutes the main focus of the present work.

In principle, the process $g g \to H q \bar{q} $ presents infrared divergences
when the quark-antiquark pair in the final state is collectively soft,
and/or when one or both of the quarks are collinear
to the direction of an incoming gluon.
However, thanks to the factorisation properties of QCD,
in double-unresolved configurations the amplitude $A^{(0,2)}_{gg\to H q \bar{q}}$
can be approximated by universal factors times the reduced amplitude  $A^{(-2,2)}_{q\bar{q}\to H}$ (that is, of order $g^3$) which is identically zero.
Indeed, the triangle one-loop diagrams for $q \bar{q} \to H$
require a mass insertion for the chirality flip
and therefore vanishes for massless onshell quarks.
This explains why the interference
involving the amplitude $A^{(0,2)}_{gg\to Hq\bar{q}}$ only requires the subtraction of \emph{single}-unresolved infrared limits,
while the interference built upon the amplitude $A^{(-1,2)}_{gq\to Hq}$ does not require IR subtraction at all.

The same observations can be made, perhaps more intuitively,
by inspecting the representative Feynman diagrams depicted in Fig.~\ref{allEWdiagrams}.
It is straightforward to see that propagators of massless partons
which do not belong to closed loops can go on-shell
only in the graphs of type~\ref{subfig:box_div_ggHddb} and~\ref{subfig:triangle_I_ggHddb}.
In the case of the diagram~\ref{subfig:box_div_ggHddb}, this happens
when antiquark $\bar{d}_5$ becomes collinear to gluon $g_2$
such that the hard scattering subgraph corresponds to diagram~\ref{diagAewborn}.
By contrast, in the kinematic limit where
quark $d_4$ is collinear to gluon $g_1$
and quark $\bar{d}_5$ is collinear to gluon $g_2$,
both non-loop propagators of graph~\ref{subfig:triangle_I_ggHddb} are singular.
The subgraph that describes the hard scattering process, however,
evaluates to zero for massless quarks as explained before,
thus avoiding the singularity.
In the limit where only one of the quarks is collinear to an incoming gluon,
the hard part of diagram~\ref{subfig:triangle_I_ggHddb} matches that of graph~\ref{diagBewborn}.

From the observations drawn so far,
we conclude that for the local subtraction of implicit singularities
it is sufficient to consider standard \gls{NLO} initial-collinear counterterms.
These subtraction terms are to be added back, analytically integrated over
the unresolved degrees of freedom yielding explicit poles in the dimensional regulator $\epsilon$.
These poles cancel against those part of the \gls{PDF} renormalisation counterterms,
as guaranteed by collinear beam factorisation, thus rendering the complete computation finite.

The formal expression which describes this subtraction procedure
and the combination with \gls{PDF} renormalisation counterterms reads:
\begin{align}
\sigma^{(m,n)}_{g g \to H q \bar{q}} = &
\int_0^1 \dd{x_1} \int_0^1 \dd{x_2} f_g(x_1) f_g(x_2) \bigg\{
\nonumber\\&\qquad
\int \dd{\Phi_{H q \bar{q}}} \Big[
\mathcal{M}^{(m,n)}_{g g \to H q \bar{q}}
\mathbf{J}(\phi_{H q\bar{q}})
- \sum_\pi \mathbf{C}_{g q} \otimes
\mathcal{M}^{(m-1,n)}_{g q \to H q}
\mathbf{J}(\tilde{\phi}_{H q })
\Big]
\label{intreal}
\\&\qquad
+ \sum_\pi \int \dd{\Phi_{H q}}
\int_0^1 \dd{\xi}
\left[
	\intd{\mathbf{C}_{gq}}(\xi)
	+ \mathrm{\Delta}_{g q}(\xi)
\right]
\mathcal{M}^{(m-1,n)}_{g q \to H q}
\mathbf{J}(\phi_{H q})
\label{intcoll}
\bigg\},
\end{align}
where the dependences
on the factorisation and renormalisation scales $\mu_F$ and $\mu_R$
as well as on the kinematic inputs for the matrix elements
have been suppressed for brevity.
The sums run over the four permutations $\pi$
that are obtained exchanging the quark and the antiquark in the final state
and/or the two initial-state gluons among themselves.
The symbol $\mathbf{C}_{ij}$ denotes the local counterterm
for particles $i$ and $j$ going collinear
and $\intd{\mathbf{C}_{ij}}$ its counterpart analytically integrated over the unresolved degrees of freedom.
The observable functions are indicated with $\mathbf{J}$,
and $\mathrm{\Delta}_{ik}$ is the \gls{PDF} renormalisation kernel
for parton with flavour $i$ to change into species $k$ before entering the hard process.
The notation $\tilde{\phi}_{Hq}$ indicates
reduced kinematics of lower multiplicity
which are obtained by mapping a pair of collinear partons to a massless parent.
The concrete expressions of all subtraction ingredients
closely follow ref.~\cite{Somogyi:2009ri}
and are presented more explicitly in appendix~\ref{app:collinearlimits},
where we also explicitly show that our subtraction counterterms
correctly regulate the relevant collinear singularities.

\section{Setup of the computation and numerical results}
\label{sec:setup}

The amplitudes $A^{(0,2)}_{gg\to H q \bar{q}}$ and $A^{(-1,2)}_{gq\to H q}$
that factorise a Higgs coupling to weak bosons 
were first computed analytically (for massless quarks only) in ref.~\cite{Harlander:2008xn},
in the different context of \gls{NLO} QCD corrections to weak vector-boson fusion.
In the present case and as indicated in Table~\ref{InterferencesTable}, in order to obtain contributions
to $\sigma^{(1,1)}_{gg\to H q \bar{q}}$ and $\sigma^{(0,1)}_{gq\to H q}$,
these amplitudes must be interfered against their corresponding QCD analog.

Nowadays such one-loop amplitudes are readily available 
from many automated one-loop matrix-element generators.
However, a high degree of flexibility is necessary in order to be able 
to select the relevant diagrams and interferences,
and to construct the appropriate subtraction terms.
This motivates our choice of generating the relevant one-loop squared amplitudes using \MadLoop~\cite{Hirschi:2011pa}, 
part of \mgamclong~\cite{Alwall:2014hca} (henceforth abbreviated \mgamc), as it can efficiently generate and
interfere~\cite{Hirschi:2015iia} arbitrary one-loop amplitudes in the \gls{SM} and beyond.
\MadLoop\ uses \ninja~\cite{Peraro:2014cba,Hirschi:2016mdz} and \oneloop~\cite{vanHameren:2010cp}, or alternatively \collier~\cite{Denner:2016kdg}, 
for performing one-loop reductions and for the evaluation of the scalar one-loop master integrals.
We present in appendix~\ref{app:benchmark} some details about the generation procedure as well as
benchmark numbers in order to facilitate the reproduction of our results.
Moreover, we have cross-checked \MadLoop's numerical implementation of the amplitudes $A^{(2,0)}_{gg\to H q \bar{q}}$ and $A^{(0,2)}_{gg\to H q \bar{q}}$
against a completely independent and analytical computation described in appendix~\ref{app:analyticalamplitude}.

As already mentioned, we choose to renormalise all unstable particles in the complex-mass scheme~\cite{Denner:1991kt,Frederix:2018nkq} and consider the \gls{SM} input parameters given in Table~\ref{tableParams}.

\begin{table}[h!]
\begin{center}
{\setlength\doublerulesep{1.5pt}   
 \aboverulesep=0ex 
 \belowrulesep=-0.5ex
\begin{tabular}{ll|ll|ll}

\toprule[1pt]\midrule[0.3pt]
	Parameter &  \multicolumn{1}{l}{value} & Parameter & \multicolumn{1}{l}{value} & Parameter & value 
\\	\hline	\midrule
	\gls{PDF} set & {\tt PDF4LHC15\_nlo\_30} & $\mu_R=\mu_F$ & $M_H/2, M_H$ & $M_t$ & 174.3
\\
	$\alpha_{S}(m_Z^2)$ & from \gls{PDF} set & $G_F$ & $\frac{\pi \alpha}{\sqrt{2}m_W^2(1-m_W^2/m_Z^2)}$ & $\Gamma_{t}$ & 1.35408
\\
	$\sqrt{\hat{s}}$ & 13000 & $\alpha^{-1}$ & 132.507 & $\frac{y_{t} v }{ \sqrt{2}}$ & $m_t$ 
\\
	$\bar{M}_Z$ & 91.188 & $\bar{\Gamma}_{Z}$ & 2.42823 & $M_{b}$ & 0.0
\\
	$\bar{M}_{W}$ & 80.419 & $\bar{\Gamma}_{W}$ & 2.02844 & $\frac{y_{b} v }{ \sqrt{2}}$ & 0.0
\\
	$M_H$ & 125.0  & $\Gamma_{H}$ & 0.0 & $V^{CKM}_{ij}$ & $\delta_{ij}$
\\
\midrule[0.3pt]\bottomrule[1pt]
\end{tabular}
}
\end{center}
\caption{\label{tableParams} \gls{SM} parameters used for obtaining all numerical results presented in Table~\ref{mainresults}. Dimensionful parameters are given in GeV. Lower-case mass parameters correspond to their complex-valued counterpart in the complex-mass scheme, \emph{i.e.} $m_W=\sqrt{\bar{M}_W^2 - i \bar{\Gamma}_{W} \bar{M}_W}$.}
\end{table}

The numerical Monte-Carlo integration as well as the necessary IR subtraction procedure, presented in Eqs.~\ref{intreal} and~\ref{intcoll} as well as in appendix~\ref{app:collinearlimits}, have been implemented in a private extension of \mgamc\ currently under development.
The poles in the dimensional regulator $\epsilon$ have been checked to cancel as expected.
\footnote{This check of course only considers the convoluted term of Eq.~\ref{intcoll} as our computation involves no virtual contribution. Also, for the pole cancellation to occur, it is important to restrict the initial state contributions to gluons only, as poles from the beam factorisation terms $\mathrm{\Delta}_{q g}$ and $\mathrm{\Delta}_{q q}$ remain uncanceled given that we ignore the corresponding real-emission subprocesses.}
Moreover, we have validated our code by comparing \gls{NLO} QCD cross sections
against results from \mgamc\ for the processes $p p \to Z$ and $p p \to H$,
the latter in the Higgs Effective Theory.

\renewcommand{\arraystretch}{1.05}
\begin{table}[hbtp]
\begin{center}
{\setlength\doublerulesep{1.5pt}   
	\aboverulesep=0.1ex 
	\belowrulesep=-0.11ex
\begin{tabular}{cr@{\hspace{6pt}}c@{\hspace{6pt}}l|cr@{\hspace{6pt}}c@{\hspace{6pt}}l}
	
\toprule[1pt]\midrule[0.3pt]
 cross section &\multicolumn{3}{c}{ \multirow{ 2}{*}{[fb]}} & cross section & \multicolumn{3}{c}{ \multirow{ 2}{*}{[fb]}} \\
interferences &\multicolumn{3}{c}{ }&  squared amplitudes & \multicolumn{3}{c}{ }
	\\\midrule

\multirow{ 2}{*}{$\sigma^{(\alpha_s^3\alpha^2)}_{gg\to Hq\bar{q}}$}
& $11.93$ & $\pm$ & $0.04$
& \multirow{ 2}{*}{$\sigma^{(\alpha_s^2\alpha^3,\textrm{no-HZ})}_{gg\to Hq\bar{q}}$}
& $-0.260$ & $\pm$ & $0.004$ \\ 
& $13.31$ & $\pm$ & $0.08$ &
& $-2.135$ & $\pm$ & $0.003$ \\ \cdashlinelr{2-4} \cdashlinelr{6-8}

\multirow{ 2}{*}{$\sigma^{(\alpha_s^3\alpha^2)}_{gg\to Hb\bar{b}}$}
& $-5.94$ & $\pm$ & $0.03$
& \multirow{ 2}{*}{$\sigma^{(\alpha_s^2\alpha^3,\textrm{no-HZ})}_{gg\to Hb\bar{b}}$}
& $3.867$ & $\pm$ & $ 0.008$ \\
& $-7.36$ & $\pm$ & $0.03$ &
& $0.882$ & $\pm$ & $0.006$ \\ \cdashlinelr{2-4} \cdashlinelr{6-8}

\multirow{ 2}{*}{$\sigma^{(\alpha_s^2\alpha^2)}_{qg\to Hq}$ + $\sigma^{(\alpha_s^2\alpha^2)}_{\bar{q}g\to H\bar{q}}$}
& $-163.9\phantom{0}$ & $\pm$ & $0.1$
& \multirow{ 2}{*}{$\sigma^{(\alpha_s\alpha^3)}_{qg\to Hq}$ + $\sigma^{(\alpha_s\alpha^3)}_{\bar{q}g\to H\bar{q}}$}
& $52.3\phantom{00}$ & $\pm$ & $0.2$ \\
& $-137.0\phantom{0}$ & $\pm$ & $0.2$ &
& $48.6\phantom{00}$ & $\pm$ & $0.1$ \\ \cdashlinelr{2-4} \cdashlinelr{6-8}

\multirow{ 2}{*}{$\sigma^{(\alpha_s^2\alpha^2)}_{bg\to Hb}$ + $\sigma^{(\alpha_s^2\alpha^2)}_{\bar{b}g\to H\bar{b}}$}
& $20.95$ & $\pm$ & $0.04$
& \multirow{ 2}{*}{$\sigma^{(\alpha_s\alpha^3)}_{bg\to Hb}$ + $\sigma^{(\alpha_s\alpha^3)}_{\bar{b}g\to H\bar{b}}$}
& $13.78\phantom{0}$ & $\pm$ & $0.05$ \\
& $19.45$ & $\pm$ & $0.06$ &
& $13.82\phantom{0}$ & $\pm$ & $0.02$ \\ \cline{1-4} \cdashlinelr{6-8}

\multirow{ 2}{*}{$ \sigma_{\textrm{total}}^{\textrm{interf.+squared}}$}
& $30.9\phantom{0}$ & $\pm$ & $0.2$
& \multirow{ 2}{*}{$\sigma^{(\alpha_s^2\alpha^2,\bar{\Gamma}_Z=0)}_{gg\to HZ}$}
& $98.17\phantom{0}$ & $\pm$ & $ 0.05$ \\
& $24.9\phantom{0}$ & $\pm$ & $0.2$ &
& $76.27\phantom{0}$ & $\pm$ & $0.03$ \\

\midrule
\multicolumn{8}{c}{ \multirow{ 2}{*}{ $p_T(H)>400$ GeV } }  \\
\\
\midrule
\multirow{ 2}{*}{$\sigma^{(\alpha_s^3\alpha^2)}_{gg\to Hq\bar{q}}$}
& $-0.0054\phantom{0}$ & $\pm$ & $0.0002$
& \multirow{ 2}{*}{$\sigma^{(\alpha_s^2\alpha^3,\textrm{no-HZ})}_{gg\to Hq\bar{q}}$}
& $0.00390$ & $\pm$ & $0.00003$ \\
& $0.00674$ & $\pm$ & $0.00008$ &
& $0.00154$ & $\pm$ & $0.00004$ \\ \cdashlinelr{2-4} \cdashlinelr{6-8}

\multirow{ 2}{*}{$\sigma^{(\alpha_s^3\alpha^2)}_{gg\to Hb\bar{b}}$}
& $-0.0093\phantom{0}$ & $\pm$ & $0.0002$
& \multirow{ 2}{*}{$\sigma^{(\alpha_s^2\alpha^3,\textrm{no-HZ})}_{gg\to Hb\bar{b}}$}
& $0.0363\phantom{0}$ & $\pm$ & $ 0.0003$ \\
& $-0.00197$ & $\pm$ & $0.00009$ &
& $0.0118\phantom{0}$ & $\pm$ & $0.0002$ \\ \cdashlinelr{2-4} \cdashlinelr{6-8}

\multirow{ 2}{*}{$\sigma^{(\alpha_s^2\alpha^2)}_{qg\to qH}$ + $\sigma^{(\alpha_s^2\alpha^2)}_{\bar{q}g\to H\bar{q}}$}
& $-1.005\phantom{00}$ & $\pm$ & $0.003$
& \multirow{ 2}{*}{$\sigma^{(\alpha_s\alpha^3)}_{qg\to Hq}$ + $\sigma^{(\alpha_s\alpha^3)}_{\bar{q}g\to H\bar{q}}$}
& $0.1019\phantom{0}$ & $\pm$ & $ 0.0002$ \\
& $-0.7486\phantom{0}$ & $\pm$ & $0.0005$ &
& $0.0841\phantom{0}$ & $\pm$ & $0.0001$ \\ \cdashlinelr{2-4} \cdashlinelr{6-8}

\multirow{ 2}{*}{$\sigma^{(\alpha_s^2\alpha^2)}_{bg\to Hb}$ + $\sigma^{(\alpha_s^2\alpha^2)}_{\bar{b}g\to H\bar{b}}$}
& $-0.0326\phantom{0}$ & $\pm$ & $0.0001$
& \multirow{ 2}{*}{$\sigma^{(\alpha_s\alpha^3)}_{bg\to Hb}$ + $\sigma^{(\alpha_s\alpha^3)}_{\bar{b}g\to H\bar{b}}$}&
$0.1033\phantom{0}$ & $\pm$ & $0.0003$ \\
& $-0.0268\phantom{0}$ & $\pm$ & $0.00003$ &
& $0.0950\phantom{0}$ & $\pm$ & $0.0002$ \\ \cline{1-4} \cdashlinelr{6-8}

\multirow{ 2}{*}{$ \sigma_{\textrm{total}}^{\textrm{interf.+squared}}$}
& $-0.502\phantom{00}$ & $\pm$ & $0.003 $
& \multirow{ 2}{*}{$\sigma^{(\alpha_s^2\alpha^2,\bar{\Gamma}_Z=0)}_{gg\to HZ}$}
& $0.3049\phantom{0}$ & $\pm$ & $0.0006$\\
& $-0.3615\phantom{0}$ & $\pm$ & $0.0008$ &
& $ 0.2159\phantom{0}$ & $\pm$ & $0.0003$
\\\midrule
\end{tabular}
}
\end{center}
\caption{\label{mainresults} Fully and semi inclusive cross sections obtained with \gls{SM} input parameters given in Table~\ref{tableParams} for the processes $gg\to Hq\bar{q}$ and $gg\to Hb\bar{b}$, as (partial) contributions to the corrections of order $\order{\alpha_s^3\alpha^2}$ and $\order{\alpha_s^2\alpha^3}$ to inclusive Higgs production.
For the contributions of order $\order{\alpha_s^2\alpha^3}$ labelled ``no-HZ'', the diagrams of the class~\ref{subfig:triang_associated_prod_ggHddb} and~\ref{subfig:box_associated_prod_ggHddb} are ignored, as they are best accounted for in the narrow-width approximation as the \gls{LO} contribution to the process $g g \to H Z$, which is also shown.
We also report the $\order{\alpha_s^2\alpha^2}$ and $\order{\alpha_s\alpha^3}$ contributions from the quark-initiated processes $q g \to H q $ and $b g \to H b$.
Finally, we consider the boosted regime, in which the Higgs transverse momentum is required to be at least $400$ GeV. In each bracket separated by a dashed line, the upper number corresponds to the scale choice $\mu_R=\mu_F=m_H/2$ while the lower one corresponds to $\mu_R=\mu_F=m_H$.}
\end{table}

Our results are presented in Table~\ref{mainresults}.
Along with the different contributions to the inclusive cross section for Higgs production,
we also report the semi-inclusive cross sections for the production of a Higgs boson with transverse momentum larger than 400 GeV.
The motivation to consider this boosted Higgs regime is twofold.
On one side, it mimics typical experimental selection cuts used to reduce backgrounds and study new physics effect prominent in that regime.
On the other side, it selects a region of phase space where real emissions are typically hard
and the relative importance of the corrections computed in this work may in principle be enhanced.

We find that the squared contributions of order $\order{\alpha_s^2\alpha^3}$ can be suppressed compared to their $\order{\alpha_s^3\alpha^2}$ counterpart by less than what is expected by their parametric ratio $\alpha/\alpha_s$.
This is for example the case for the processes involving $b$ quarks, and it can be explained by the kinematic suppressions interfering contributions are typically subject to.

Also, contributions of order $\order{\alpha_s^m\alpha^n}$ with $m+n=4$ are numerically more relevant than those with $m+n=5$ in spite of their suppression by one quark luminosity.
The quark-initiated \gls{EW} corrections are however still small in comparison with the whole $\sigma^{(\alpha_s^3\alpha^2)}_{pp\to H+X}$, and can thus be safely neglected as already observed in  ref.~\cite{Keung:2009bs}.
These two observations reinforce the conclusion that the contributions to $\sigma^{(\alpha_s^3\alpha^2)}_{gg\to Hq\bar{q}}$ that are computed in this work and which have been neglected up to this point are of similar (ir)relevance to that of other neglected terms of weak origin.

The cross section $\sigma^{(\alpha_s^3\alpha^2)}_{gg\to Hb\bar{b}}$ from only final-state $b$ quarks reveals that contributions featuring Higgs production from the internal top quark line (see Fig.~\ref{diagF}) are comparable and of opposite sign to that of emissions from internal weak bosons.
This fact contrasts with the study of ref.~\cite{Degrassi:2004mx} of the two-loop amplitude $A^{(0,2)}_{gg\to H}$ where it was instead found that Higgs emissions from internal top quarks only contribute to less than 2\% of the complete amplitude at this order, and could thus be safely ignored in the computation of the three-loop amplitude $A^{(2,2)}_{gg\to H}$ of refs.~\cite{Bonetti:2017ovy,Bonetti:2018ukf,Anastasiou:2018adr}.
Indeed, the higher partonic collision energy probed by $A^{(1,2)}_{gg\to Hb\bar{b}}$ enhances contributions from internal top-quark Higgs emissions, even more so in the boosted regime.
Similarly, the same mechanism enables the bottom-quark initiated contribution $\sigma^{(\alpha_s^2\alpha^2)}_{bg\to Hb}$ at the same level as that of the channels initiated by each other valence quark flavour.

The squared contribution $\sigma^{(\alpha_s^2\alpha^3,\textrm{no-HZ})}_{gg\to Hb\bar{b}}$ omits the diagrams~\ref{subfig:triang_associated_prod_ggHddb} and~\ref{subfig:box_associated_prod_ggHddb} featuring a $Z$ boson decay since it is best accounted for in the narrow-width approximation.
It is however clear that the extent to which one should consider Higgs production in association with an on-shell $Z$ boson depends on the particular observable considered.
We chose to report here the quantity $\sigma^{(\alpha_s^2\alpha^2,\bar{\Gamma}_Z=0)}_{gg\to HZ}$ only to serve as an upper bound to this contribution.

Notice that the contribution $\sigma^{(\alpha_s^2\alpha^3,\textrm{no-HZ})}_{gg\to Hq\bar{q}}$ is negative, despite involving squared amplitudes.
This originates from the finite logarithms in the \gls{PDF} renormalisation term $\mathrm{\Delta}_{gq}$ and integrated counterterm $\intd{\mathbf{C}_{gq}}$, stemming from dimensional regularisation.
Our results also highlight that considering a subset of higher-order corrections and factorising only a particular combination of initial-state flavours typically yields a large dependence on the factorisation scale.
This is especially true for squared amplitude contributions and the boosted regime, for which the chosen fixed scales proportional to the Higgs mass (as it is tailored to the prediction of the inclusive Higgs production cross section) are not well suited in light of the significantly larger collision energies probed.
We choose to report here \emph{absolute} factorisation scale dependency, given that some contributions can be accidentally close to 
zero\footnote{This is for example the case in $\sigma^{(\alpha_s^2\alpha^3,\textrm{no-ZH})}_{gg\to Hb\bar{b}}|_{\mu_F=m_H}$ where Higgs emissions from weak bosons are close to equal and opposite in sign to emissions from internal top-quarks, and in $\sigma^{(\alpha_s^2\alpha^3,\textrm{no-ZH})}_{gg\to Hq\bar{q}}|_{\mu_F=m_H/2}$ where the cancellation occurs between the hard reals and the logarithms in $\xi$ part of the integrated counterterms $\intd{\mathbf{C}_{gq}}$. }
for one of the two scale choices.
A more detailed analysis of the sensitivity of these contributions to the factorisation scale is beyond the scope of this work.

The overall magnitude of all contributions computed here is such that they can be safely neglected in light of the total size of mixed QCD-\gls{EW} corrections, which is estimated to be of the order of $2$ pb with an associated uncertainty in the range of $200$ fb \cite{Anastasiou:2018adr,Bonetti:2018ukf}.
The aggregated sum $\sigma_{\textrm{total}}$ of all contributions computed in this work is only meant to serve as qualitative highlight of that fact.
Our results then further support the factorisation approximation when accounting for mixed \gls{EW} and QCD corrections to inclusive Higgs production.

The hierarchy of the various terms is altered when considering the boosted Higgs regime,
where the kinematic suppression of gluon-initiated interference contributions is strong enough to make them of the same order or smaller than their squared counterpart.
All interference contributions also become negative in this case, while the square term 
$\sigma^{(\alpha_s^2\alpha^3,\textrm{no-ZH})}_{gg\to Hq\bar{q}}$ 
is now positive as hard real emissions become dominant.
Overall, none of the contributions computed plays a significant role in that scenario either, given that the pure QCD contribution is estimated in ref.~\cite{Chen:2016zka} to be 25 fb with a large theoretical uncertainty exceeding 20\%.

\section{Conclusion}

The large QCD corrections to inclusive Higgs production at LHC13 calls for accounting for mixed \gls{EW} and QCD corrections in a multiplicative scheme, that is assuming their complete factorisation.
In light of the accuracy sought-after for this process, it is important to assess the validity of this factorisation assumption by explicitly computing $\sigma^{(1,1)}_{pp\to H+X}$, namely the mixed QCD and \gls{EW} correction of order $\mathcal{O}(\alpha_s^3\alpha^2)$ to the Higgs inclusive cross section.

To this end, two groups~\cite{Bonetti:2017ovy,Bonetti:2018ukf,Anastasiou:2018adr} computed $\sigma^{(1,1)}_{gg\to H+X}$ and found that it supports the hypothesis that \gls{EW} corrections factorise. These works however neglected the quark-initiated components as well as the ``double-real'' channel $g g \to H q \bar{q}$ and we confirm here that these terms can be safely neglected, amounting to about $5\%$ of the total mixed \gls{EW} and QCD corrections.
We verified that our conclusions also apply when imposing that the Higgs transverse momentum lies above $400$ GeV.
The interference nature of the contributions $\sigma^{(1,1)}_{gg\to H q\bar{q}}$ and $\sigma^{(1,1)}_{gq\to H q}$ renders them prone to kinematic suppressions, and we indeed found that the square of the one-loop \gls{EW} amplitudes involved can be larger than naively expected from their parametric suppression of $\alpha/\alpha_s$.
The selective nature of the contributions computed in this work is such that they feature a large factorisation scale dependency, further stressing that their inclusion would require to also consider \emph{all} other partonic channels.

Besides further establishing the validity of the hypothesis assumed when accounting for weak corrections to inclusive Higgs production, our work also showcases the novel flexibility brought by recent developments in the realm of automated one-loop matrix element generation and Monte-Carlo integration for higher-order computations.

\section*{Acknowledgments}

We are greatly indebted to
C.~Anastasiou,
V.~del~Duca,
N.~Deutschmann,
E.~Furlan,
B.~Mistlberger
and C.~Specchia
for fruitful discussions.
This project has received funding
from the European Research Council (ERC)
under grant agreements No 772099 (JetDynamics) and No 694712 (PertQCD).
We thank the Galileo Galilei Institute for Theoretical Physics
for the hospitality
and the INFN for partial support during the completion of this work.

\newpage
\begin{appendices}
\numberwithin{equation}{section}

\section{Initial-collinear counterterms}
\label{app:collinearlimits}

In this appendix, we detail all ingredients that are necessary
for the subtraction of implicit singularities
outlined in Eqs.~\ref{intreal}--\ref{intcoll},
and we demonstrate that the matrix element for $g_1 g_2 \to q_3 \bar{q}_4 H_5$
is correctly regulated in the regions of phase space
close to unresolved configurations.
As already announced in Section~\ref{sec:singularities},
the construction and notation follow closely ref.~\cite{Somogyi:2009ri}.

Let us begin with the expression of the local initial-collinear counterterms
in Eq.~\ref{intreal}.
In general, in order for these counterterms to approximate the matrix element
point by point in the phase space,
spin correlations need to be taken into account.
Suppressing the coupling orders, we define
\begin{multline}
\label{eq:Cgq}
	[\mathbf{C}_{g_1q_3} \otimes
	\mathcal{M}_{\bar{q}_{13} g_2 \to \bar{q}_4 H_5}]
	(\phi_{q_3 \bar{q}_4 H_5})
	\equiv (8\pi\alpha_s\mu^{2\epsilon})
	\frac{1}{s_{g_1 q_3}} \hat{P}_{g_1q_3}^{ss'}(1/z)
	\frac{\omega(\bar{q}_{13})}{\omega(g_1)}
	\\\times
	\mathcal{M}_{\bar{q}_{13} g_2 \to \bar{q}_4 H_5}^{ss'}
	(\tilde\phi_{\bar{q}_4 H_5})
	\theta(y_0-y),
\end{multline}
where $s$ and $s'$ respectively specify the spin of the quark
which enters the reduced amplitude and the corresponding conjugate.
The factor $\omega(\bar{q})/\omega(g)$ accounts for the different averaging
on the initial state spins and colours in the matrix elements
and equals $N_c/(N_c^2-1)$ in four spacetime dimensions.
The symbol $\hat{P}_{gq}^{ss'}$ denotes the final-final $qg$ splitting function
\begin{equation}
\label{eq:PgqFS}
	\hat{P}_{gq}^{ss'}(z)
	\equiv
	\delta^{ss'} C_F \left[ \frac{1+(1-z)^2}{z} - \epsilon z \right],
\end{equation}
and the variable $z$ in Eq.~\ref{eq:Cgq} is computed using
\begin{equation}
	z \equiv \frac{Q\cdot (p_1-p_3)}{Q\cdot p_1},
\end{equation}
with $Q = p_1+p_2$.
For the process at hand, since the parton which enters the hard process
after the splitting is always a quark,
spin correlations are absent as indicated by $\delta^{ss'}$
in Eq.~\ref{eq:PgqFS}.
The momentum mapping that we use to determine
the reduced phase-space point $\tilde\phi_{qH}$
is the one used for two initial-state partons
in Catani--Seymour dipole subtraction
(see Section 5.5 of \cite{Catani:1996vz}).
\footnote{
Note that this mapping involves recoiling against all final state particles,
and it would not be efficient for studying differential Higgs observables.
}
Finally, the Heaviside $\theta$ function at the end of Eq.~\ref{eq:Cgq}
controls the region of phase-space where the counterterm is active
through the parameter $y_0$,
which determines the range for the variable $y \equiv 2p_1\cdot p_3 / Q^2$. 

At this point, all the elements needed to check that Eq.~\eqref{intreal}
only features integrable singularities have been presented.
In order to validate our subtraction
and assess the numerical stability of the integrand
which is built from the interference of two one-loop amplitudes,
we start from a random resolved kinematic configuration
and examine the behaviour as different collinear limits are approached.
We control the distance from any given unresolved limit using
a scaling variable $\lambda$,
which is engineered to approach the singular configuration
at a pace such that the phase-space volume
between $\lambda$ and $\lambda+\dd{\lambda}$
is proportional to $\lambda$ itself.
For a kinematic configuration with a centre-of-mass energy of 1 TeV and 
in the case of a collinear pair, the typical invariant mass
is then $\order{\text{1 GeV}}$ for $\lambda=10^{-6}$
and $\order{\text{1 MeV}}$ for $\lambda=10^{-12}$.
Under the same conditions and in the case of two collinear pairs, 
the typical invariant mass of each of them is $\order{\text{1 GeV}}$ for $\lambda=10^{-12}$
and $\order{\text{1 MeV}}$ for $\lambda=10^{-24}$ instead.
For the sake of concreteness,
we consider the partonic subprocess $g_1 g_2 \to b_3 \bar{b}_4 H_5$,
with the understanding that all qualitative features are identical
in the case of light quark flavours in the final state.

\begin{figure}[h]
\centering
\begin{subfigure}[b]{\linewidth}
	\includegraphics[width=0.5\linewidth,trim=0.5cm 0.2cm 0.5cm 0cm,clip]{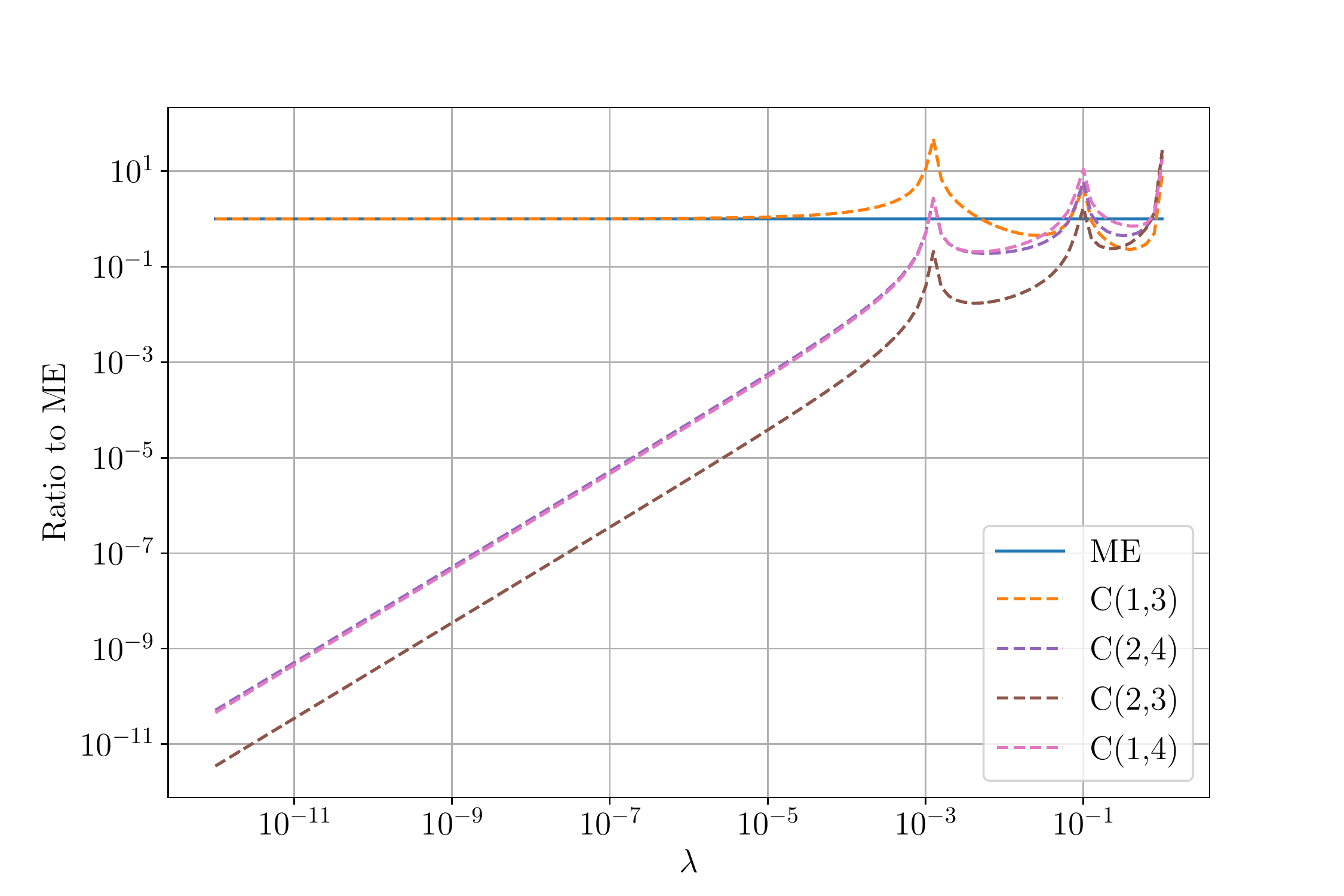}
	\includegraphics[width=0.5\linewidth,trim=0.5cm 0.2cm 0.5cm 0cm,clip]{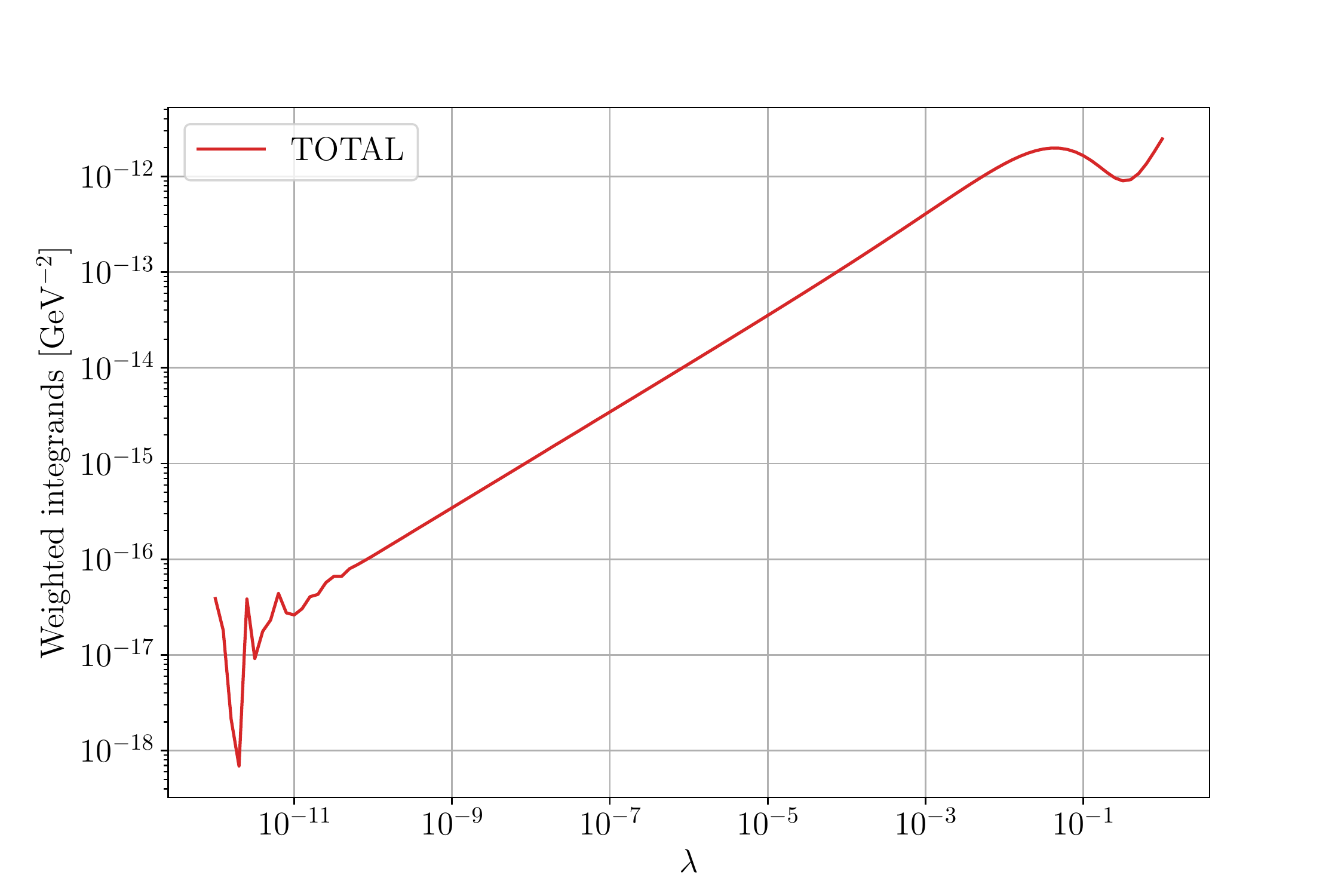}
	\caption{$C(1,3)$ limit}
	\label{c13dp}
\end{subfigure}
\begin{subfigure}[b]{\linewidth}
\includegraphics[width=0.5\linewidth,trim=0.5cm 0.2cm 0.5cm 0cm,clip]{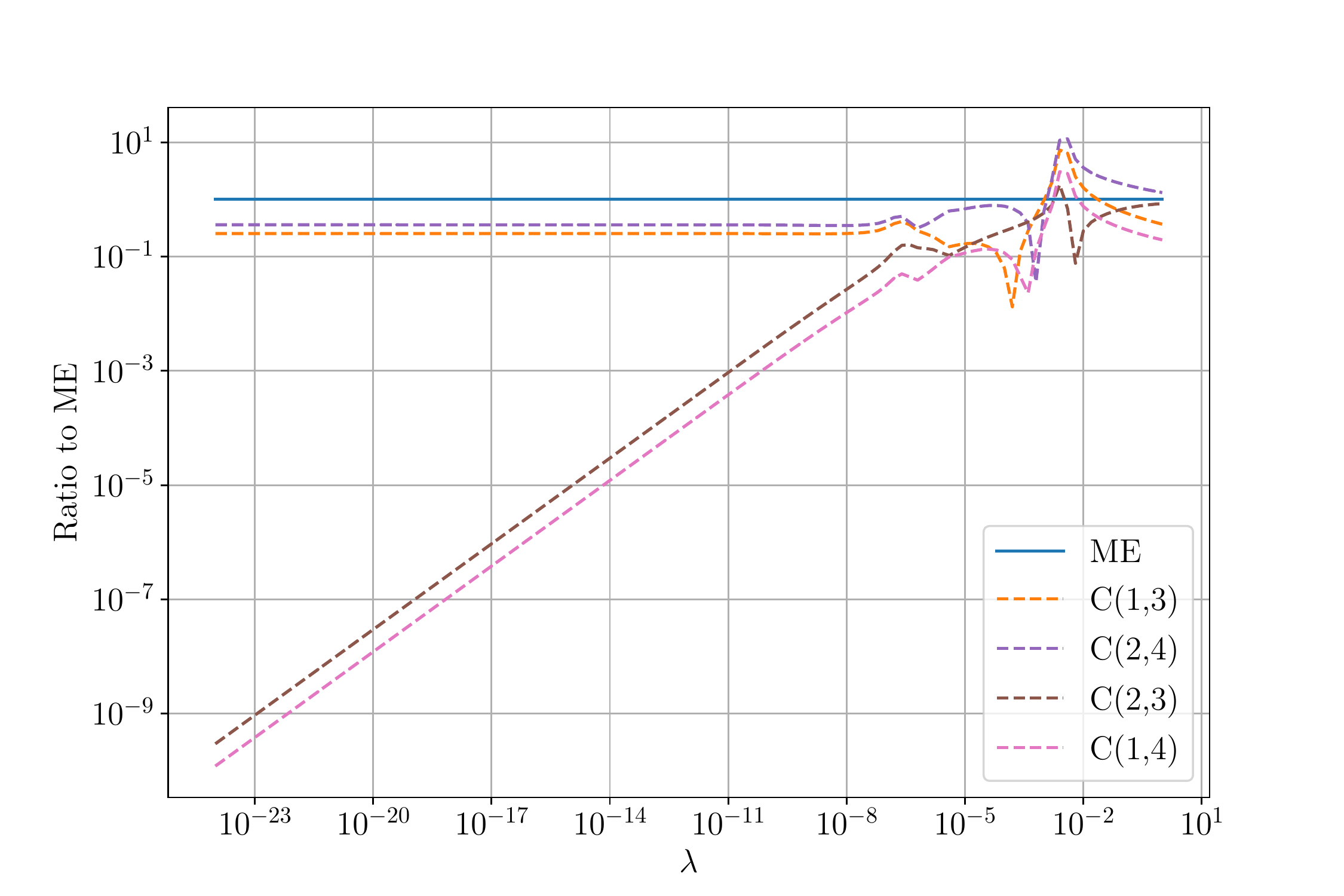}
\includegraphics[width=0.5\linewidth,trim=0.5cm 0.2cm 0.5cm 0cm,clip]{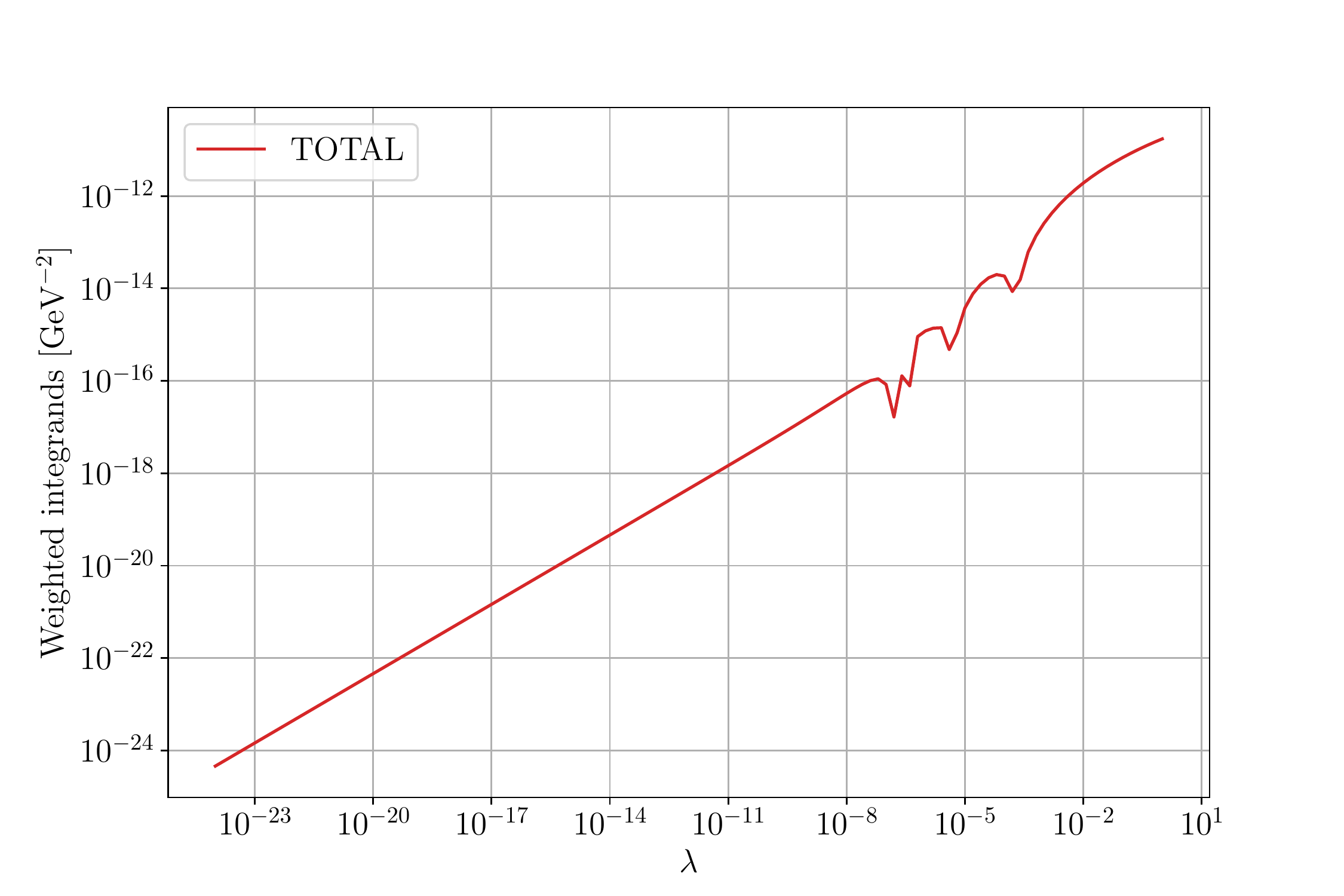}
\caption{$C(1,3)C(2,4)$ limit}
\label{c13c24dp}
\end{subfigure}
\begin{subfigure}[b]{\linewidth}
\includegraphics[width=0.5\linewidth,trim=0.5cm 0.2cm 0.5cm 0cm,clip]{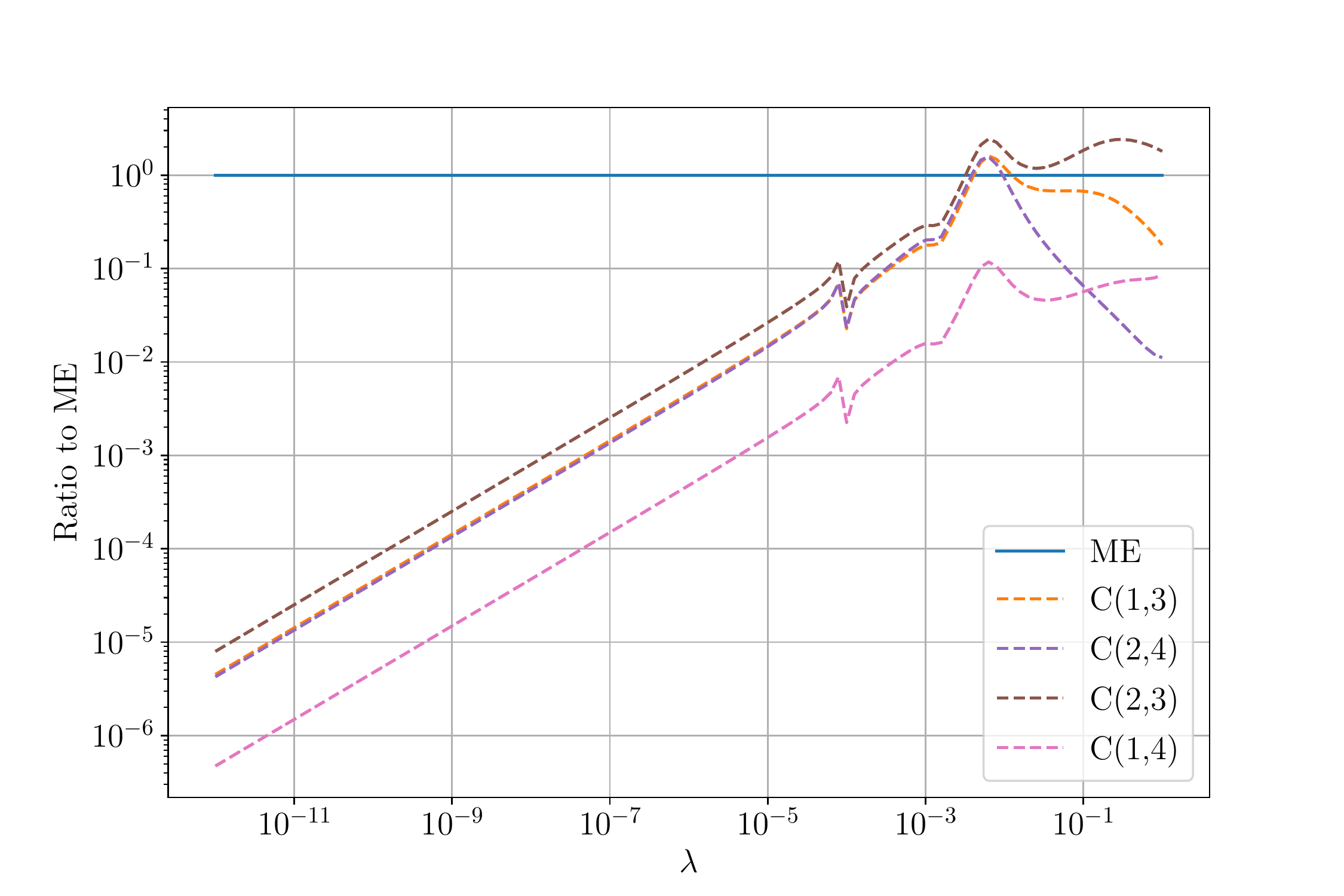}
\includegraphics[width=0.5\linewidth,trim=0.5cm 0.2cm 0.5cm 0cm,clip]{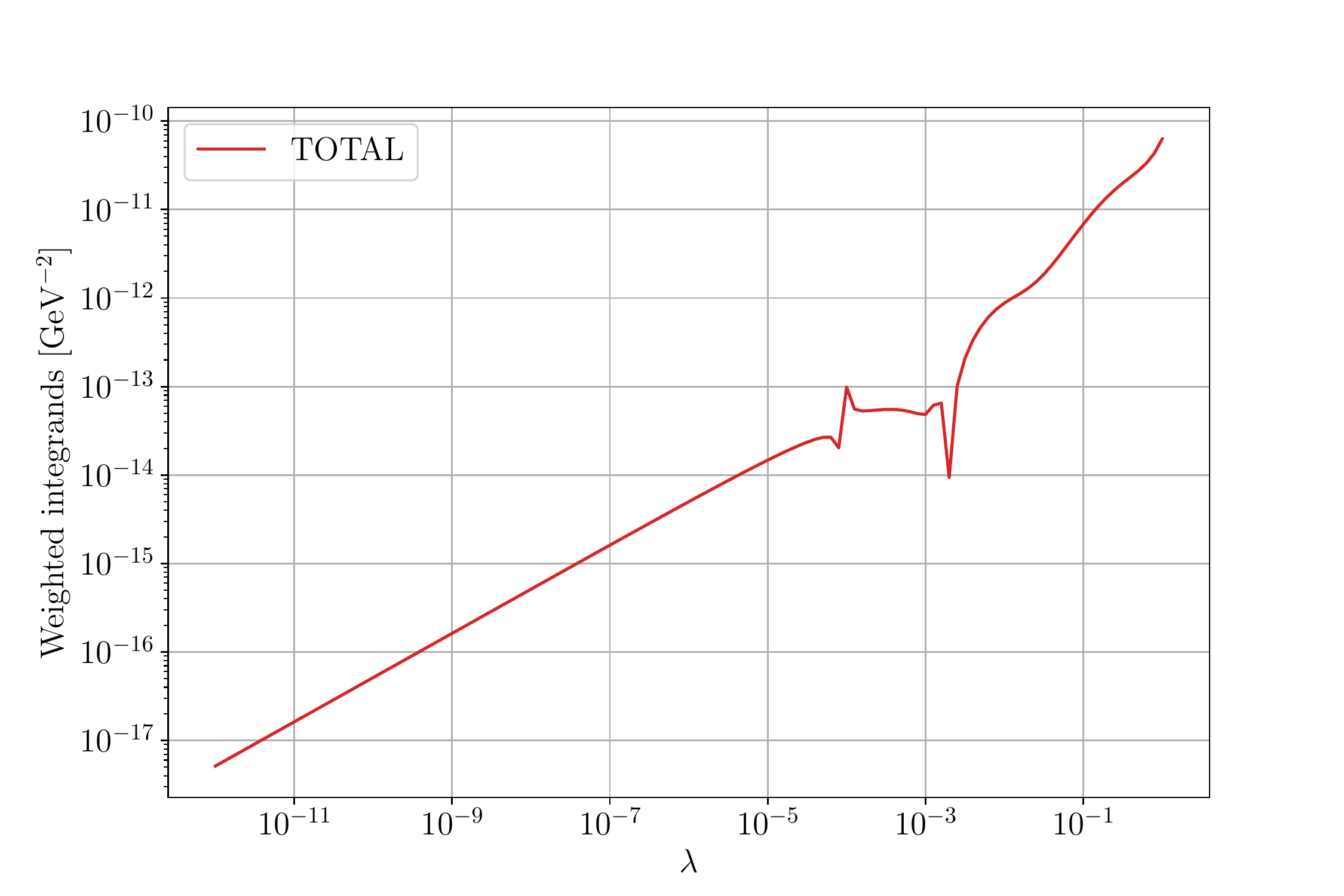}
\caption{$C(3,4)$ limit}
\label{c34dp}
\end{subfigure}
\caption{
Behaviour of the terms in Eq.~\ref{intreal}
for the process $g_1 g_2 \to b_3 \bar{b}_4 H_5$
when approaching different unresolved limits.
See text for details.
}
\label{fig:testIRlimits}
\end{figure}

In Fig.~\ref{fig:testIRlimits}, we display the behaviour
of the matrix element interference and its four initial-collinear counterterms
as a function of $\lambda$ for a given starting kinematic configuration.
The left panels simply show the ratio of counterterms to the matrix element.
In the right panels, we plot their sum weighted by $\lambda$,
which is representative of the contribution to the total integral
coming from a neighbourhood of $\lambda$.
We therefore expect the integral to be convergent
if this quantity tends to zero when $\lambda \to 0$.
In Fig.~\ref{c13dp} we consider the limit $C(1,3)$,
where the matrix element is approximated by the counterterm $C(1,3)$
and all other terms in the sum over $\pi$ of Eq.~\eqref{intreal} are regular.
The cases of $C(2,4)$, $C(1,4)$ and $C(2,3)$ are fully analogous.
In Fig.~\ref{c13c24dp}
we study the limit of two collinear pairs $C(1,3)C(2,4)$ which,
as discussed in Section~\ref{sec:singularities},
does not require any additional treatment
since the matrix element for $q \bar{q} \to H$
at order $\order{g_s g^2}$ is zero.
Finally, in Fig.~\ref{c34dp} we consider the limit $C(3,4)$
to confirm that the matrix element of order $\order{\alpha_s^2\alpha^2}$
for $g g \to b \bar{b} H$ does not feature a non-integrable divergence
when the two quarks in the final state are collinear.
We note that the figures in this section
can be sensitive to the numerical stability parameters of \MadLoop\ 
which, among other things, control when to switch
to a slower quadruple precision evaluation.
Further discussion of this technical aspect
is however beyond the scope of this work and we limit ourselves to reporting here that 
all Monte-Carlo integrations performed in this work could be successfully
carried out using \MadLoop's default parameters\footnote{We note however that it proved to be necessary to employ an estimate of \MadLoop's accuracy based on the comparison of two separate numerical evaluations that differ by a Lorentz transformation of the kinematic inputs (by setting \MadLoop's parameter {\tt NRotations\_DP} to 1).}.
Incidentally, we observe that in order to obtain results at the level of precision
needed for this work it is not necessary to introduce a technical cutoff.

The integral of the collinear counterterm over the unresolved phase space
has been computed in \cite{Somogyi:2009ri} and reads
\begin{multline}
	\intd{\mathbf{C}_{gq}}(\xi) =
	\frac{\alpha_s}{2\pi} S_\epsilon
	\left[\frac{\mu_R^2}{Q^2}\right]^\epsilon
	\frac{T_R}{C_F} \bigg\{
		[\xi^2+(1-\xi)^2] \bigg[ -\frac{1}{\epsilon}
		+ \ln(1-\xi)  (1+\theta[\xi-(1-y_0)])
		\\
		+ \ln(y_0)\theta[(1-y_0)-\xi]
		\bigg]
		+2\xi(1-\xi)
	\bigg\}
	+\order{\epsilon},
\end{multline}
where $Q = p_1+p_2$ for the \emph{reduced} process $g_1 q_2 \to q_3 H_4$
and we have defined
\begin{equation}
	S_\epsilon \equiv \frac{(4\pi)^\epsilon}{\Gamma(1-\epsilon)}.
\end{equation}
The explicit pole in the dimensional regulator $\epsilon$ 
featured by this integrated counterterm
is cancelled by the contribution from \gls{PDF} renormalisation,
which is given by
\begin{equation}
	\mathrm{\Delta}_{gq}(\xi)
	=
	\frac{\alpha_s}{2\pi} S_\epsilon \frac{1}{\epsilon}
	\left[\frac{\mu_R^2}{\mu_F^2}\right]^{\epsilon}
	P_{gq}(\xi),
\end{equation}
where the relevant Altarelli--Parisi splitting kernel reads
\footnote{
Note that in our notation the first subscript indicates
the parton extracted from the hadron according to its \gls{PDF},
and the second one denotes the parton that enters the hard process.
}
\begin{equation}
	P_{gq}(\xi) = T_R [\xi^2+(1-\xi)^2].
\end{equation}
We have confirmed that in our implementation of this subtraction scheme
the sum of \eqref{intreal} and \eqref{intcoll} does not depend on $y_0$,
which provides a non-trivial cross-check of $\order{\epsilon^0}$ terms.

\section{Analytical computation of the amplitudes $A^{(2,0)}_{gg\to H q \bar{q}}$ and $A^{(0,2)}_{gg\to H q \bar{q}}$}
\label{app:analyticalamplitude}

The analytic validation is performed by computing the form factors for the QCD background depicted in the diagrams Fig.~\ref{allQCDdiagrams} and the \gls{EW} contributions with sample diagrams shown in Figs.~\ref{subfig:pentagon_ggHddb} to \ref{subfig:triangle_I_ggHddb}.
While we retain the full quark mass dependence for the QCD background,
we assume massless quarks for the \gls{EW} contributions.
The computation is performed in $D=4-2\epsilon$ dimensions.
However, due to the special reduction of the scalar pentagon integrals, the final result is only valid for the provided order $\order{\epsilon^0}$ in the dimensional regulator (see Sec.~\ref{subsec:integrals_and_eval}).

The amplitude for the \gls{EW} process $g_1 g_2\to q_3 \bar{q}_4 H_5$ may be written in the general form
\begin{align}
	A^{(0,2)}_{g g \to q \bar{q} H}
	&=A^{(0,2),\vcr\vcr}_{g g \to q \bar{q} H}+A^{(0,2),\axl\vcr}_{g g \to q \bar{q} H}+A^{(0,2),\vcr\axl}_{g g \to q \bar{q} H}+A^{(0,2),\axl\axl}_{g g \to q \bar{q} H} \\
	&=\varepsilon_{\mu_1}(p_1)\varepsilon_{\mu_2}(p_2)\bar{u}^{s_3}(p_3)v^{s_4}(p_4)\mathcal{F}_{s_3 s_4}^{\mu_1 \mu_2}, 
	\label{eq:formfactor_generic}
\end{align}
where in the following we will denote the form factor by $\mathcal{F}^{\mu_1 \mu_2}$, suppressing the spinor indices.
We separate couplings of the quarks to the \gls{EW} gauge bosons according to
\begin{align}
	& u d W^+\propto g_\vcr\gamma^\mu + g_\axl \gamma^\mu \gamma^5, \qquad
	u d W^-\propto g_\vcr^* \gamma^\mu + g_\axl^* \gamma^\mu \gamma^5 \quad \text{and} \\
	& q q Z\propto g_{\vcr,Z}  \gamma^\mu+g_{\axl,Z} \gamma^\mu \gamma^5,
\end{align}
and refer to $g_\vcr$ as the vector and to $g_\axl$ as the axial coupling constant.
In the following, we restrict ourselves to the case of $d$ quarks in the final state for concreteness.
We will first discuss the computation of $A^{(0,2),\vcr\vcr}_{g g \to d \bar{d} H} \propto |g_\vcr|^2$ and then argue that this piece is sufficient to determine the complete amplitude.

 In order to compute the form factors, we first generate all contributing diagrams with \QGraf\ \cite{Nogueira:1991ex} and perform the color-, Dirac- and Lorentz algebra in Mathematica, whereas the $\gamma$ traces are performed using \FORM\ \cite{Vermaseren:2000nd}.
 For more compact expressions, we choose an axial gauge for the external gluons $g_1$ and $g_2$, such that
\begin{align}
	p_1\cdot \varepsilon_1(p_1)=0 && p_2\cdot \varepsilon_1(p_1)=0 && p_2\cdot \varepsilon_2(p_2)=0 && p_1 \cdot \varepsilon_1(p_2)=0,
\end{align}
with the physical polarization sum
\begin{align}
	\sum \limits_{\text{polarization}}\varepsilon^{*}_{\mu}(p_i)\varepsilon^{}_{\nu}(p_i)
	=-g_{\mu \nu}+\frac{p_{1\mu} p_{2\nu}+p_{2\mu} p_{1\nu}}{p_1 \cdot p_2} \qquad \text{for } i=1,2. 
\end{align}
As a direct consequence of the gauge choice, terms in the form factors proportional to $p_1^{\mu_1}$, $p_2^{\mu_1}$, $p_1^{\mu_2}$ or $p_2^{\mu_2}$ can be set to zero, since they will not contribute to the amplitude.
Internal gauge bosons and quarks are treated in the Feynman gauge.

\subsection{Tensor reduction to scalar integrals}

The form factor $\mathcal{F}^{\mu_1\mu_2}$ can be written as
\begin{align}
	\mathcal{F}^{\mu_1\mu_2}
	=\sum_i \alpha_i \mathcal{S}^{\mu_1\mu_2}_i
	=\sum_i \alpha_i \int \frac{N^{\mu_1 \mu_2}(k)}{D_1\dots D_{m_i}} \dd[D]{k},
\end{align}
where the $\mathcal{S}^{\mu_1\mu_2}_i$ denote tensor integrals.
We can reduce the tensor integrals $\mathcal{S}^{\mu_1\mu_2}_i$ to scalar integrals $S_k$:
\begin{align}
	\mathcal{S}^{\mu_1\mu_2}_i=\sum_j T_j^{\mu_1\mu_2}(p_1,p_2,p_3,p_4)S_j.
\end{align}
To achieve the above decomposition, we use a particular flavour of Passarino--Veltman tensor reduction \cite{Passarino:1978jh}.
The reduction of the tensor integrals is discussed here by writing only the integrand numerators $N^{\mu_1\mu_2}(k)$, keeping in mind that the identities exclusively hold at the integral level.

As a first step we strip off the external Lorentz structures factorising the loop momentum $N^{\mu_1\mu_2}(k)$ and write
 \begin{align}
 	N^{\mu_1\mu_2}(k)=
 	c_0^{\mu_1\mu_2}
 	+c_1^{\mu_1 \mu_2; \alpha_1} k_{\alpha_1}
 	+c_2^{\mu_1 \mu_2; \alpha_1 \alpha_2} k_{\alpha_1}k_{\alpha_2}
 	+c_3^{\mu_1 \mu_2; \alpha_1 \alpha_2 \alpha_3} k_{\alpha_1}k_{\alpha_2}k_{\alpha_3}+ \dotsc, 
 	\label{eq:general_form_tensor_numerator}
 \end{align}
 where the tensor coefficients $c_i$ only involve $\gamma$-matrices, external momenta $p_i$ and the metric tensor $g$. The tensor reduction is performed with the fully symmetric tensor numerators $\tilde{N}^{(\alpha_1\alpha_2\dots\alpha_n)}(k)=k^{\alpha_1}k^{\alpha_2}\cdots k^{\alpha_n}$. Performing the loop integration of the tensor integral
\begin{align}
\mathcal{S}^{(\alpha_1\alpha_2\dots\alpha_n)}_i
=\int \frac{\tilde{N}^{(\alpha_1\alpha_2\dots\alpha_n)}(k)}{D_1\dots D_{m_i}} \dd[D]{k}
=\sum_j t_j^{(\alpha_1\alpha_2\dots\alpha_n)}(g,p_1,p_2,p_3,p_4)c_j,
\end{align}
will result in Lorentz tensors $t_j^{(\alpha_1\alpha_2\dots\alpha_n)}$ which are also completely symmetric in the internal Lorentz indices $\alpha_i$. The symmetric tensor $t^{(\alpha_1\dots \alpha_n)}$ is given by
\begin{align}
	t^{(\alpha_1\dots \alpha_n)}=\frac{1}{n!}\sum_{\sigma\in \Sigma_n}t^{\alpha_{\sigma(1)} \dots \alpha_{\sigma(n)}},
\end{align}
where $\Sigma_n$ is the symmetric group of order $n$, e.g.
\begin{align}
	t_{p_1,p_2}^{(\alpha_1\alpha_2)}=p_1^{(\alpha_1} p_2^{\alpha_2)} 
	=\frac{1}{2}\left(p_1^{\alpha_1}p_2^{\alpha_2}+p_1^{\alpha_2}p_2^{\alpha_1}\right). 
\end{align}
Reduction with respect to a fully symmetric tensor basis reduces the number of tensor structures to be considered in the Ansatz significantly, while still remaining completely algorithmic.

A further simplification arises from the fact that the only underlying scalar topology depending on the full external kinematics $p_1,p_2,p_3,p_4$ is the pentagon displayed in Fig.~\ref{subfig:pentagon_ggHddb}.
Every other diagram will yield scalar integrals with reducible external kinematics yielding results depending on a reduced set of Mandelstam variables only.
In general we group all diagrams into families characterized by their minimal set of external momenta and perform the tensor reduction for the tensor numerators $\tilde{N}^{(\alpha_1\alpha_2\dots\alpha_n)}(k)$ in each family separately with respect to the reduced external kinematics.
This approach keeps the intermediate expressions obtained from tensor- and \gls{IBP} reduction very compact.
The biggest matrix we have to invert is a $24\times24$ matrix for the rank $3$ Lorentz tensor integrals of the pentagon diagram.
To perform the analytic matrix inversion we employ the computer algebra system \Fermat\ \cite{FermatCAS}, which takes below a minute on one core of a modern computer.
In order to obtain the form factor in terms of scalar integrals we then insert the solutions back into \eqref{eq:general_form_tensor_numerator}.

With the tensor decomposition described, we are able to express the form factors for the QCD background and the vector-vector part of the \gls{EW} contributions in terms of the following 20 tensor structures:
\begin{gather}
\begin{aligned}
	T_{1}^{\mu_1\mu_2}&=\slashed{p} {}_{1}^{{}_{}}  \gamma^{\mu_1}  \gamma^{\mu_2},&
	T_{2}^{\mu_1\mu_2}&=\slashed{p} {}_{2}^{{}_{}}  \gamma^{\mu_1}  \gamma^{\mu_2},& 
	T_{3}^{\mu_1\mu_2}&=g^{\mu_1 \mu_2} \slashed{p} {}_{1}^{{}_{}},& 
	T_{4}^{\mu_1\mu_2}&=g^{\mu_1 \mu_2} \slashed{p} {}_{2}^{{}_{}},\\ 
	T_{5}^{\mu_1\mu_2}&=\gamma^{\mu_1} p_3^{\mu_2},& 
	T_{6}^{\mu_1\mu_2}&=\gamma^{\mu_1} p_4^{\mu_2},& 
	T_{7}^{\mu_1\mu_2}&=\slashed{p} {}_{1}^{{}_{}}  \slashed{p} {}_{2}^{{}_{}}  \gamma^{\mu_1} p_3^{\mu_2},& 
	T_{8}^{\mu_1\mu_2}&=\slashed{p} {}_{1}^{{}_{}}  \slashed{p} {}_{2}^{{}_{}}  \gamma^{\mu_1} p_4^{\mu_2},\\ 
	T_{9}^{\mu_1\mu_2}&=\gamma^{\mu_2} p_3^{\mu_1},& 
	T_{10}^{\mu_1\mu_2}&=\slashed{p} {}_{1}^{{}_{}}  \slashed{p} {}_{2}^{{}_{}}  \gamma^{\mu_2} p_3^{\mu_1},& 
	T_{11}^{\mu_1\mu_2}&=\gamma^{\mu_2} p_4^{\mu_1},& 
	T_{12}^{\mu_1\mu_2}&=\slashed{p} {}_{1}^{{}_{}}  \slashed{p} {}_{2}^{{}_{}}  \gamma^{\mu_2} p_4^{\mu_1},\\ 
	T_{13}^{\mu_1\mu_2}&=\slashed{p} {}_{1}^{{}_{}} p_3^{\mu_1} p_3^{\mu_2},& 
	T_{14}^{\mu_1\mu_2}&=\slashed{p} {}_{2}^{{}_{}} p_3^{\mu_1} p_3^{\mu_2},& 
	T_{15}^{\mu_1\mu_2}&=\slashed{p} {}_{1}^{{}_{}} p_3^{\mu_1} p_4^{\mu_2},& 
	T_{16}^{\mu_1\mu_2}&=\slashed{p} {}_{2}^{{}_{}} p_3^{\mu_1} p_4^{\mu_2},\\ 
	T_{17}^{\mu_1\mu_2}&=\slashed{p} {}_{1}^{{}_{}} p_3^{\mu_2} p_4^{\mu_1},& 
	T_{18}^{\mu_1\mu_2}&=\slashed{p} {}_{2}^{{}_{}} p_3^{\mu_2} p_4^{\mu_1},& 
	T_{19}^{\mu_1\mu_2}&=\slashed{p} {}_{1}^{{}_{}} p_4^{\mu_1} p_4^{\mu_2},&
	T_{20}^{\mu_1\mu_2}&=\slashed{p} {}_{2}^{{}_{}} p_{4}^{\mu_1} p_4^{\mu_2}.
\end{aligned}
\end{gather}

\subsection{Evaluation of scalar integrals}
\label{subsec:integrals_and_eval}

The \gls{IBP} reduction of the remaining scalar integrals is performed using the program \Kira\ \cite{Maierhoefer:2017hyi,Maierhofer:2018gpa}.
We decompose the scalar pentagon integrals appearing as master integrals following ref.~\cite{Bern:1992em}.
This decomposition relates the pentagon in $4-2\epsilon$ dimensions to a linear combination of all boxes obtainable by pinching one of the propagators, and the pentagon in $6-2\epsilon$ dimensions multiplied by a prefactor of order $\epsilon$.
Since the pentagon in six dimensions is finite, the additional term involving the six-dimensional pentagon is of order $\order{\epsilon}$ and can be omitted for the computation at hand.

We find that all form factors are finite diagram-by-diagram, but order $\order{\epsilon}$ coefficients of the bubbles appear explicitly in the final amplitude.
\footnote{This is particular to our approach and originates from the \gls{IBP} reduction of the scalar integrals.}
The relevant coefficients are given by
\begin{align}
	&b_0(s;m^2,0)=h(\epsilon)\int \frac{1}{(k^2-m^2)(k-p)^2}\dd[D]{k}
	\nonumber \\
	\label{eq:b0sm0}
	&\ =
	\left(\frac{\mu^2}{m^2}\right)^\epsilon\left[ \frac{1}{\epsilon}
		+2-\frac{(x-1) \ln(1-x)}{x} \right.\\
	& \  \left.+\left(4+\frac{\pi^2}{6}+\frac{(1-x)}{2 x}\left(2 \Li_2\left(-\frac{x}{1-x}\right)-\ln^2(1-x)+4 \ln (1-x)\right)\right)\epsilon+\order{\epsilon^2}\right],
	\nonumber \\
	\label{eq:b0s00}
	&b_0(s;0,0)=h(\epsilon)\int \frac{1}{k^2(k-p)^2}\dd[D]{k}
	=\left(-\frac{s}{\mu ^2}\right)^{-\epsilon}\left[\frac{1}{\epsilon }+2+4 \epsilon +\order{\epsilon^2}\right], \\
	\label{eq:a0m}
	&a_0(m)=h(\epsilon)\int \frac{1}{k^2-m^2}\dd[D]{k}
		=	m^2 \left(\frac{\mu ^2}{m^2}\right)^{\epsilon } \left[ \frac{1}{\epsilon }+1+\left(1+\frac{\pi ^2}{6}\right) \epsilon + \order{\epsilon^2}\right],
\end{align}
where $x=s/m^2$.
The normalization
\begin{equation}
	h(\epsilon)
	\equiv \frac{\mu^{2\epsilon}}{i \pi^{2-\epsilon}}
	\frac{\Gamma(1-2\epsilon)}{\Gamma(1-\epsilon)^2 \Gamma(\epsilon+1)},
\end{equation}
is chosen to match the convention of \oneloop\ \cite{vanHameren:2010cp},
which is also used to evaluate the remaining non-trivial scalar master integrals.

In the evaluation of the expressions above,
some care is needed in order to evaluate multi-valued functions on their physical Riemann sheet.
The convention for numerical implementations of such functions is that
the value assigned on the cut is the one
coming around the finite endpoint of the cut in a counter-clockwise direction
\cite{British2003c}. 
The Feynman prescription, however, dictates to replace $s$ with $s+i\eta$
and take the limit $\eta\downarrow 0$, which gives
\begin{equation}
	\lim_{\eta\downarrow 0} \ln\left(1-\frac{s+i\eta}{m^2}\right)=
	\begin{cases}
		\ln(1-x) &s<m^2, \\
		\ln(1-x)-2i\pi & s>m^2,
	\end{cases}
\end{equation}
if the right-hand side respects the convention.
The same holds for $s > 0$ in the expansion of the massless bubble.
It is easy to see that for the dilogarithm in \eqref{eq:b0sm0}, instead,
the physical sheet coincides with the conventional one for all $x\neq1$.

\subsection{Relations between the axial and vector parts of the amplitude}

In the previous section we discussed the computation of the vector part $A^{(0,2),\vcr\vcr}_{g g \to d \bar{d} H}$.
In what follows, we restrict the discussion to a single quark family with a diagonal CKM matrix, $g_\vcr=g_\vcr^*$ and $g_\axl=g_\axl^*$.
The generalisation to all families of light quarks is straightforward and purely combinatorial.
Since there are no closed fermion loops, we do not have to worry about ambiguous traces of $\gamma^5$ in $4-2\epsilon$ dimensions and we may take the $D$-dimensional $\gamma^5$ to be anticommuting.
\footnote{
Our choice corresponds to the NDR treatment of $\gamma^5$ (see e.g.\ \cite{Jegerlehner:2000dz}).
Note however, that the amplitude is finite diagram by diagram and traces over $\gamma^5$ enter only in the interference of the AV-part with e.g.\ the QCD background.
Since for the interference there is no explicit $\epsilon$-dependence anymore, the traces can be treated as four-dimensional objects, without the need of imposing additional constraints.
}

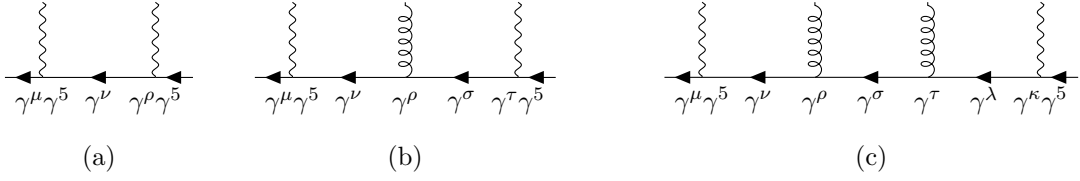
\begin{figure}[h!]
\begin{center}
\begin{subfigure}{0.2\textwidth}
\centering
\begin{tikzpicture}[scale=0.9, every node/.style={scale=0.9}]
	\begin{feynman}
		\vertex (a1);
		\vertex[right=0.5cm of a1] (a2) ;
		\vertex[right=1.5cm of a2] (a3) ;
		\vertex[right=0.5cm of a3] (a4);
		\vertex[above=1cm of a2] (b2);
		\vertex[above=1cm of a3] (b3);
		\vertex[below=0.01cm of a2] (l1) {\($$\gamma^{\mu}\gamma^5$$\)};
		\vertex[below=0.01cm of a3] (l2) {\($$\gamma^{\rho}\gamma^5$$\)}; 
		\diagram* {
			{[edges=fermion]
				(a4) -- (a3),
				(a2) -- (a1),
			},
			(a3) -- [fermion, edge label=\($$\gamma^{\nu\vphantom{5}}$$\)] (a2),
			(a2) -- [boson] (b2),
			(a3) -- [boson] (b3),
		};
	\end{feynman}
\end{tikzpicture}
\caption{}\label{subfig:internalGammaStrucA}
\end{subfigure} \hfill
\begin{subfigure}{0.3\textwidth}
\centering
\begin{tikzpicture}[scale=0.9, every node/.style={scale=0.9}]
	\begin{feynman}
		\vertex (a1);
		\vertex[right=0.5cm of a1] (a2) ;
		\vertex[right=1.5cm of a2] (a3) ;
		\vertex[right=1.5cm of a3] (a4) ;
		\vertex[right=0.5cm of a4] (a5) ;
		\vertex[above=1cm of a2] (b2);
		\vertex[above=1cm of a4] (b4);
		\vertex[above=1cm of a3] (g3);
		\vertex[below=0.01cm of a2] (l1) {\($$\gamma^{\mu}\gamma^5$$\)};
		\vertex[below=0.01cm of a3] (l2) {\($$\gamma^{\rho\vphantom{5}}$$\)}; 
		\vertex[below=0.01cm of a4] (l3) {\($$\gamma^{\tau}\gamma^5$$\)}; 
		\diagram* {
			{[edges=fermion]
				(a5) -- (a4),
				(a2) -- (a1),
			},
			(a3) -- [fermion, edge label=\($$\gamma^{\nu\vphantom{5}}$$\)] (a2),
			(a4) -- [fermion, edge label=\($$\gamma^{\sigma\vphantom{5}}$$\)] (a3),
			(a2) -- [boson] (b2),
			(a4) -- [boson] (b4),
			(a3) -- [gluon] (g3),
		};
	\end{feynman}
\end{tikzpicture}
\caption{}\label{subfig:internalGammaStrucB}
\end{subfigure}
\begin{subfigure}{0.48\textwidth}
\centering
\begin{tikzpicture}[scale=0.9, every node/.style={scale=0.9}]
	\begin{feynman}
		\vertex (a1);
		\vertex[right=0.5cm of a1] (a2) ;
		\vertex[right=1.5cm of a2] (a3) ;
		\vertex[right=1.5cm of a3] (a4) ;
		\vertex[right=1.5cm of a4] (a5) ;
		\vertex[right=0.5cm of a5] (a6) ;
		\vertex[above=1cm of a2] (b2);
		\vertex[above=1cm of a5] (b5);
		\vertex[above=1cm of a3] (g3);
		\vertex[above=1cm of a4] (g4);
		\vertex[below=0.01cm of a2] (l1) {\($$\gamma^{\mu}\gamma^5$$\)};
		\vertex[below=0.01cm of a3] (l2) {\($$\gamma^{\rho\vphantom{5}}$$\)}; 
		\vertex[below=0.01cm of a4] (l3) {\($$\gamma^{\tau\vphantom{5}}$$\)}; 
		\vertex[below=0.01cm of a5] (l4) {\($$\gamma^{\kappa}\gamma^5$$\)}; 
		\diagram* {
			{[edges=fermion]
				(a6) -- (a5),
				(a2) -- (a1),
			},
			(a3) -- [fermion, edge label=\($$\gamma^{\nu\vphantom{5}}$$\)] (a2),
			(a4) -- [fermion, edge label=\($$\gamma^{\sigma\vphantom{5}}$$\)] (a3),
			(a5) -- [fermion, edge label=\($$\gamma^{\lambda}$$\)] (a4),
			(a2) -- [boson] (b2),
			(a5) -- [boson] (b5),
			(a3) -- [gluon] (g3),
			(a4) -- [gluon] (g4),
		};
	\end{feynman}
\end{tikzpicture}	
\caption{}\label{subfig:internalGammaStrucC}
\end{subfigure}
\caption{The relevant $\gamma$ matrix structures for $A^{(0,2),\axl\axl}_{g g \to d \bar{d} H}$. Fig.~\ref{subfig:internalGammaStrucA} corresponds to triangle diagrams, Fig.~\ref{subfig:internalGammaStrucB} corresponds to box diagrams and Fig.~\ref{subfig:internalGammaStrucC} corresponds to pentagon diagrams.}
\label{fig:internalGammaStrucAA}
\end{center}
\end{figure}

\begin{figure}[h!]
\begin{subfigure}{0.25\textwidth}
\centering
	\begin{tikzpicture}[scale=0.9, every node/.style={scale=0.9}]
		\begin{feynman}
		\vertex (a1);
		\vertex[right=1cm of a1] (a2) ;
		\vertex[right=1cm of a2] (a3) ;
		\vertex[below=0.2cm of a1] (p1);
		\vertex[right=0.7cm of p1] (p2) ;	
		\vertex[above=1cm of a2] (b2);
		\vertex[below=0.01cm of a2] (l1) {\($$\gamma^{\rho\vphantom{5}}\gamma^5$$\)}; 
		\diagram* {
			{[edges=fermion]
				(a3) -- (a2) --(a1),
			},
			(p2) -- [fermion, edge label=\($$p_3$$\)] (p1),
			(a2) -- [boson] (b2),
		};
	\end{feynman}
	\end{tikzpicture}
\caption{}\label{subfig:externalGammaStrucA}
\end{subfigure} \hfill
\begin{subfigure}{0.25\textwidth}
\centering
\begin{tikzpicture}[scale=0.9, every node/.style={scale=0.9}]
	\begin{feynman}
		\vertex (a1);
		\vertex[right=1cm of a1] (a2) ;
		\vertex[right=1.5cm of a2] (a3) ;
		\vertex[right=0.5cm of a3] (a4);
		\vertex[below=0.2cm of a1] (p1);
		\vertex[right=0.7cm of p1] (p2) ;	
		\vertex[above=1cm of a3] (b3);
		\vertex[above=1cm of a2] (b2);
		\vertex[below=0.01cm of a2] (l1) {\($$\gamma^{\mu\vphantom{5}}$$\)};
		\vertex[below=0.01cm of a3] (l2) {\($$\gamma^{\rho}\gamma^5$$\)}; 
		\diagram* {
			{[edges=fermion]
				(a4) -- (a3),
				(a2) -- (a1),
			},
			(p2) -- [fermion, edge label=\($$p_3$$\)] (p1),
			(a3) -- [fermion, edge label=\($$\gamma^{\nu\vphantom{5}}$$\)] (a2),
			(a2) -- [gluon] (b2),
			(a3) -- [boson] (b3),
		};
	\end{feynman}
\end{tikzpicture}
\caption{}\label{subfig:externalGammaStrucB}
\end{subfigure} \hfill
\begin{subfigure}{0.48\textwidth}
\centering
	\begin{tikzpicture}[scale=0.9, every node/.style={scale=0.9}]
	\begin{feynman}
		\vertex (a2);
		\vertex[right=1cm of a2] (a3) ;
		\vertex[right=1.5cm of a3] (a4) ;
		\vertex[right=1.5cm of a4] (a5) ;
		\vertex[right=0.5cm of a5] (a6) ;
		\vertex[below=0.2cm of a2] (p1);
		\vertex[right=0.7cm of p1] (p2) ;	
		\vertex[above=1cm of a5] (b5);
		\vertex[above=1cm of a3] (g3);
		\vertex[above=1cm of a4] (g4);
		\vertex[below=0.01cm of a3] (l2) {\($$\gamma^{\rho\vphantom{5}}$$\)}; 
		\vertex[below=0.01cm of a4] (l3) {\($$\gamma^{\tau\vphantom{5}}$$\)}; 
		\vertex[below=0.01cm of a5] (l4) {\($$\gamma^{\kappa}\gamma^5$$\)}; 
		\diagram* {
			{[edges=fermion]
				(a6) -- (a5),
				(a3) -- (a2),
			},
			(a4) -- [fermion, edge label=\($$\gamma^{\sigma\vphantom{5}}$$\)] (a3),
			(a5) -- [fermion, edge label=\($$\gamma^{\lambda\vphantom{5}}$$\)] (a4),
			(a5) -- [boson] (b5),
			(a3) -- [gluon] (g3),
			(a4) -- [gluon] (g4),
			(p2) -- [fermion, edge label=\($$p_3$$\)] (p1),
		};
	\end{feynman}
\end{tikzpicture}	
\caption{}\label{subfig:externalGammaStrucC}
\end{subfigure}
\caption{The relevant $\gamma$ matrix structures for $A^{(0,2),\axl\vcr}_{g g \to d \bar{d} H}$.}
\label{fig:externalGammaStrucAA}
\end{figure}

Within the purely axial amplitude $A^{(0,2),\axl\axl}_{g g \to d \bar{d} H}$ both \gls{EW} couplings are $\propto g_\axl \gamma^\mu\gamma^5$.
The $\gamma$ chains that appear in the amplitude are shown in Fig.~\ref{fig:internalGammaStrucAA}.
One always needs to do an even number of anticommutations to arrive at $\gamma^5\gamma^5=\mathds{1}$ from which immediately follows that
\begin{align}
	A^{(0,2),\axl\axl}_{g g \to d \bar{d} H}
	=\frac{|g_\axl|^2}{|g_\vcr|^2}A^{(0,2),\vcr\vcr}_{g g \to d \bar{d} H}.
\end{align} 
The axial-vector piece $A^{(0,2),\axl\vcr}_{g g \to d \bar{d} H}$ features the $\gamma$ chains shown in Fig.~\ref{fig:externalGammaStrucAA}.
These chains represent the cases where only the vertex closest to the outgoing $d$ quark (of momentum $p_3$) contributes with an axial coupling.
\footnote{The case where only the vertex ``furthest'' to the outgoing $d$ quark contributes with the axial coupling is completely analogous.}
It is easy to see that an uneven number of anticommutations is needed to bring $\gamma^5$ to the beginning of every spinor chain appearing in the process.
The form factor for the $\axl\vcr$ part of the amplitude is therefore given by
\begin{align}
	\mathcal{F}^{\mu\nu}_{\axl\vcr} =
	-2 \frac{g_\axl}{g_\vcr} \gamma^5 \mathcal{F}^{\mu\nu}_{\vcr\vcr}.
\end{align}
We thus conclude that the complete \gls{EW} amplitude can be determined from its purely vector piece.

\subsection{Ancillary files}

The notation employed for the ancillary files is the following:
We write every form factor in the ancillary files as the scalar product
\begin{align}
\mathcal{F}_{s_1 s_2,lm}^{\mu\nu,ab} = (T_{s_1 s_2,lm}^{\mu\nu,ab})_i S_i,
\end{align}
where the vector $T$ spans the direct product of colour and Lorentz spaces.
The pairs of indices $a, b$ and $l, m$ are associated with the adjoint and fundamental representations of ${\rm SU}(3)$ respectively;
$\mu, \nu$ are the Lorentz indices and $s_1, s_2$ are the spinor ones.

The interference between two amplitudes $A$ and $\tilde{A}$ in this notation then reads
\begin{equation}
\mathcal{M} = 2\frac{1}{4}\frac{1}{(N_c^2-1)^2}\mathrm{Re}\left(\tilde{s}^*_i B_{ij} s_j\right),
\end{equation}
where $B$ is the \emph{structure matrix} obtained
summing over all colours, spins and polarisations:
\begin{equation}
B_{ij} = \sum
\varepsilon_{\mu}(p_1)\varepsilon_{\nu}(p_2)
\varepsilon^*_{\mu'}(p_1)\varepsilon^*_{\nu'}(p_2)
u^{s_1'}(p_3)\bar{v}^{s_2'}(p_4)\bar{u}^{s_1}(p_3)v^{s_2}(p_4)
\times(\tilde{T}_{s_1' s_2'}^{\mu' \nu'})_i(T_{s_1 s_2}^{\mu \nu})_j.
\end{equation}
The ancillary files contain the vector $T_{s_1 s_2,lm}^{\mu \nu,ab}$ and the vector $S$ for the QCD background, the $\vcr\vcr$ and the $\axl\vcr$ part of the \gls{EW} amplitude.
We furthermore provide the structure matrices $B$ for
$A^{(0,2),\vcr\vcr,(\axl\axl)}_{g g \to d \bar{d} H}\mathcal{A^*}^{(2,0)}_{g g \to d \bar{d} H}$
and
$A^{(0,2),\axl\vcr,(\vcr\axl)}_{g g \to d \bar{d} H}\mathcal{A^*}^{(2,0)}_{g g \to d \bar{d} H}$,
which are sufficient to reproduce analytically the one-loop mixed QCD-\gls{EW} matrix element for light quarks (excluding Higgs-strahlung contributions).

\section{Validation material}
\label{app:benchmark}

In order to facilitate the reproduction of our results, we provide below the numerical result for the matrix element $\mathcal{M}^{(\alpha_s^3\alpha^2)}_{gg\to Hd\bar{d}}$ and $\mathcal{M}^{(\alpha_s^3\alpha^2)}_{gg\to Hb\bar{b}}$ summed (averaged) over final (initial) state helicity and colour configurations for the following two kinematic points and $\alpha_s=0.118$ (other SM parameters set to the values indicated in Table~\ref{tableParams}, unless otherwise stated).
\begin{footnotesize}
\begin{figure}[h!]
\begin{adjustwidth}{-1.5cm}{-1.5cm}
\begin{subfigure}[b]{\linewidth}
  \begin{align*}
    [\textrm{GeV}]         &\phantom{=\textrm{( 0}}E&&           \phantom{\textrm{,0}}p_x                            &&\phantom{\textrm{,0}}p_y                           &&\phantom{\textrm{,0}}p_z                           &&\\    
    p_{g_1}            &=\textrm{( 500}&&                           \textrm{, \phantom{-}0}                               &&\textrm{, \phantom{-}0}                               &&\textrm{, \phantom{-}500}                           &&\textrm{)}\\    
    p_{g_2}            &=\textrm{( 500}&&                           \textrm{, \phantom{-}0}                               &&\textrm{, \phantom{-}0}                               &&\textrm{, -500}                                            &&\textrm{)}\\
    p_{h_3}            &=\textrm{( 467.7884686370085}&&\textrm{, \phantom{-}166.5707878773001}&&\textrm{, \phantom{-}373.1956790038965}&&\textrm{, -190.2109596961058}                  &&\textrm{)}\\    
    p_{d_4}            &=\textrm{( 357.8737762854649}&&\textrm{, -18.01807463012543}&&\textrm{, -341.7897831227270}&&\textrm{, \phantom{-}104.5405801225597}                  &&\textrm{)}\\
    p_{\bar{d}_5}            &=\textrm{( 174.3377550775266}&&\textrm{, -148.5527132471747}&&\textrm{, -31.40589588116942}&&\textrm{, \phantom{-}85.67037957354616}                  &&\textrm{)}
  \end{align*}
\caption{First kinematic configuration}\label{pointa}
\end{subfigure}
\begin{subfigure}[b]{\linewidth}
  \begin{align*}
    [\textrm{GeV}]         &\phantom{=\textrm{( 0}}E&&           \phantom{\textrm{,0}}p_x                            &&\phantom{\textrm{,0}}p_y                           &&\phantom{\textrm{,0}}p_z                           &&\\    
    p_{g_1}            &=\textrm{( 500}&&                           \textrm{, \phantom{-}0}                               &&\textrm{, \phantom{-}0}                               &&\textrm{, \phantom{-}500}                           &&\textrm{)}\\    
    p_{g_2}            &=\textrm{( 500}&&                           \textrm{, \phantom{-}0}                               &&\textrm{, \phantom{-}0}                               &&\textrm{, -500}                                            &&\textrm{)}\\
    p_{h_3}            &=\textrm{( 503.1176012750793}&&\textrm{, \phantom{-}183.7772678439759}&&\textrm{, \phantom{-}314.6404088273092}&&\textrm{, -323.6196064356687}                  &&\textrm{)}\\    
    p_{d_4}            &=\textrm{( 101.0581181325984}&&\textrm{, -69.50635454208810}&&\textrm{, -42.77041343901509}&&\textrm{, \phantom{-}59.60118835247730}                  &&\textrm{)}\\
    p_{\bar{d}_5}            &=\textrm{( 395.8242805923223}&&\textrm{, -114.2709133018878}&&\textrm{, -271.8699953882941}&&\textrm{, \phantom{-}264.0184180831914}                  &&\textrm{)}
  \end{align*}
\caption{Second kinematic configuration}\label{pointb}
\end{subfigure}
\end{adjustwidth}
\caption{\label{kinematicpoints}The two kinematic configurations used for the evaluation of the $\order{\alpha_s^3 \alpha^2}$ contribution to the process $g g \to H d \bar{d}$ presented in Table~\ref{benchmarkevals}.}
\end{figure}
\end{footnotesize}

The matrix elements computed are free of any explicit IR or UV divergence, so that the specific $\epsilon$-dependent normalisation factor considered in \MadLoop's conventions is irrelevant in this case. For the two kinematic points shown in Table~\ref{kinematicpoints}, we find:

\begin{table}[h!]
\begin{center}
\begin{tabular}{lcc}\midrule\midrule
 $[\text{GeV}^{-2}]$ & $\phi=\phi_{\ref{pointa}}$ & $\phi=\phi_{\ref{pointb}}$
  \\\midrule
	$\mathcal{M}^{(\alpha_s^3\alpha^2)}_{g g \to H d \bar{d}}$ & \texttt{\phantom{-}1.473268137642022e-11} & \texttt{-3.202714028092470e-09} \\
	$\mathcal{M}^{(\alpha_s^3\alpha^2)}_{g g \to H b \bar{b}}$ & \texttt{-2.120437436454854e-09}  & \texttt{-5.094650485339200e-09} \\
	\midrule
	$\mathcal{M}^{(\alpha_s^3\alpha^2,\Gamma_{t,W^\pm}=0,W^{\pm}@[\ref{subfig:pentagon_ggHddb},\ref{subfig:box_div_ggHddb},\ref{subfig:triangle_I_ggHddb}],\vcr\vcr)}_{g g \to H d \bar{d}}$ 
	& \texttt{\phantom{-}1.046690169966104e-11} & \texttt{\phantom{-}1.051226540819620e-10}\\
	\text{Evaluation of analytic result}  & 
	\texttt{\phantom{-}1.046690169966233e-11} & \texttt{\phantom{-}1.051226540819659e-10}\\ 
	\cdashlinelr{2-3}
	$\mathcal{M}^{(\alpha_s^3\alpha^2,\Gamma_{t,W^\pm}=0,W^{\pm}@[\ref{subfig:pentagon_ggHddb},\ref{subfig:box_div_ggHddb},\ref{subfig:triangle_I_ggHddb}],\axl\vcr+\vcr\axl)}_{g g \to H d \bar{d}}$ 
	& \texttt{-4.013450438936635e-11} & \texttt{-4.984414054112152e-10}\\
	\text{Evaluation of analytic result} & 
 	\texttt{-4.013450438936742e-11} & \texttt{-4.984414054111984e-10}\\
 	\cdashlinelr{2-3}
	$\mathcal{M}^{(\alpha_s^3\alpha^2,\Gamma_{t,W^\pm}=0,W^{\pm}@[\ref{subfig:pentagon_ggHddb},\ref{subfig:box_div_ggHddb},\ref{subfig:triangle_I_ggHddb}],\axl\axl)}_{g g \to H d \bar{d}}$ 
	& \texttt{\phantom{-}1.046690169966104e-11} & \texttt{\phantom{-}1.051226540819620e-10}\\
	\text{Evaluation of analytic result} &
	\texttt{\phantom{-}1.046690169966233e-11} & \texttt{\phantom{-}1.051226540819659e-10}\\
	\midrule
	$\mathcal{M}^{(\alpha_s^3\alpha^2,\Gamma_{t,Z}=0,Z@[\ref{subfig:pentagon_ggHddb},\ref{subfig:box_div_ggHddb},\ref{subfig:triangle_I_ggHddb}],\vcr\vcr)}_{g g \to H d \bar{d}}$ 
	& \texttt{\phantom{-}2.656838076288246e-12} & \texttt{\phantom{-}3.508375650188969e-11}\\
	\text{Evaluation of analytic result} &
	\texttt{\phantom{-}2.656838076288616e-12} & \texttt{\phantom{-}3.508375650189406e-11}\\
	\cdashlinelr{2-3}
	$\mathcal{M}^{(\alpha_s^3\alpha^2,\Gamma_{t,Z}=0,Z@[\ref{subfig:pentagon_ggHddb},\ref{subfig:box_div_ggHddb},\ref{subfig:triangle_I_ggHddb}],\axl\vcr+\vcr\axl)}_{g g \to H d \bar{d}}$ & 
	\texttt{-1.998115098837096e-11} & \texttt{-2.730298029116885e-10}\\
	\text{Evaluation of analytic result} &
	\texttt{-1.998115098837179e-11} & \texttt{-2.730298029116787e-10}\\
	\cdashlinelr{2-3}
	$\mathcal{M}^{(\alpha_s^3\alpha^2,\Gamma_{t,Z}=0,Z@[\ref{subfig:pentagon_ggHddb},\ref{subfig:box_div_ggHddb},\ref{subfig:triangle_I_ggHddb}],\axl\axl)}_{g g \to H d \bar{d}}$ & 
	\texttt{\phantom{-}5.365688093206777e-12} & \texttt{\phantom{-}7.085433478511003e-11}\\
	\text{Evaluation of analytic result} &
	\texttt{\phantom{-}5.365688093207525e-12} & \texttt{\phantom{-}7.085433478511895e-11}
	\\\midrule\midrule
\end{tabular}
\end{center}
\caption{
\label{benchmarkevals}
Benchmark evaluations of various matrix elements comparing numerical results from \MadLoop\ against an independent analytical derivation of the amplitude, presented in appendix~\ref{app:analyticalamplitude} (see text for details).
}
\end{table}
\newpage

The first two matrix element evaluations given in Table~\ref{benchmarkevals} are exactly those used for obtaining the results of Table~\ref{mainresults}. The next six correspond to simplified setups that are only meant to ease comparisons against independent computations.
More specifically, the matrix element denoted $\mathcal{M}^{(\alpha_s^3\alpha^2,\Gamma_{t,W^\pm,Z}=0,W^{\pm}@[\ref{subfig:pentagon_ggHddb},\ref{subfig:box_div_ggHddb},\ref{subfig:triangle_I_ggHddb}],VV)}_{g g \to H d \bar{d}}$ corresponds to the case where:
\begin{itemize}
\item all widths are set to zero (then using on-shell renormalisation conditions)
\item only the diagrams from the classes \ref{subfig:pentagon_ggHddb}, \ref{subfig:box_div_ggHddb} and \ref{subfig:triangle_I_ggHddb} with a $W^{\pm}$ in the loop are kept
\item only the vector part of the two $W^{\pm}$ interactions is considered. 
\end{itemize}
The definition of the last five matrix elements of the table is fully analogous, with 'AV+VA' indicating that the amplitude includes exactly one vector-like and one axial coupling of the electroweak boson to the quarks. \\
For each matrix element we checked numerical evaluations of the analytic result for 100 phase-space points and compare them against \MadLoop\ evaluations. We found perfect agreement at the level of the $10^{\rm th}$ digit on average.

The above matrix elements can readily be generated by \MadLoop\ (from within \mgamc\ v2.6+) using commands similar\footnote{\label{mlcommanddetails}See \url{https://cp3.irmp.ucl.ac.be/projects/madgraph/wiki/MadLoopStandaloneLibrary} for instructions on how to generate the corresponding standalone library for linking against your own code.} to the following which generates the matrix element $\mathcal{M}^{(\alpha_s^3\alpha^2)}_{gg\to Hd\bar{d}}$:

\noindent\\
~\prompt\ {\tt ~set complex\_mass\_scheme True}\\
~\prompt\ {\tt ~import loop\_qcd\_qed\_sm}\\
~\prompt\ {\tt ~generate g g > h d d\~\, [virt=QCD QED] QED\^\,2==4 QCD\^\,2==6}\\
~\prompt\ {\tt ~output my\_gg\_hddx}\\
~\prompt\ {\tt ~launch -f}\\

\noindent Note that in order to select only the diagrams of the classes~\ref{subfig:pentagon_ggHddb},~\ref{subfig:box_div_ggHddb} and~\ref{subfig:triangle_I_ggHddb},
the following {\tt --loop\_filter} option
\footnote{
Also note that the rather long loop filters indicated on the command line can alternatively be specified directly in the user-function {\tt user\_filter()} of the \mgamc\ \python\ module {\tt loop\_diagram\_generation.py}.
}
can be passed to the following {\tt generate} command, yielding the matrix element $\mathcal{M}^{(\alpha_s^3\alpha^2,\Gamma_{t,W^\pm,Z}=0,Z@[\ref{subfig:pentagon_ggHddb},\ref{subfig:box_div_ggHddb},\ref{subfig:triangle_I_ggHddb}],\vcr\vcr+\axl\vcr+\vcr\axl+\axl\axl)}_{g g \to H d \bar{d}}$:

\noindent\\
{\mbox{~\prompt\ {\tt ~generate g g > h d d\~\, / w+ w- a [virt=QCD QED] QED\^\,2==4 QCD\^\,2==6}}} \\
{\tt--loop\_filter=not(23\textbackslash\textbackslash~in\textbackslash\textbackslash~struct\_pdgs\textbackslash\textbackslash~or\textbackslash\textbackslash~250\textbackslash\textbackslash~in\textbackslash\textbackslash~struct\_pdgs) }\\

\end{appendices}

\bibliographystyle{JHEP}
\bibliography{biblio}

\end{document}